\documentclass[12pt]{article}
\usepackage{latexsym}
\usepackage{amsmath}
\usepackage{amssymb}
\usepackage[dvips]{graphicx}
\usepackage{latexsym}
\usepackage{enumerate,graphicx}
\usepackage[toc,page]{appendix}
\usepackage{setspace}

\newcommand{\C}{\mathbb{C}}
\newcommand{\CP}{\mathbb{CP}}
\newcommand{\bP}{\mathbb{P}}
\newcommand{\RP}{\mathbb{RP}}

\newcommand{\R}{\mathbb{R}}

\newcommand{\Z}{\mathbb{Z}}
\newcommand{\uu}{{\bf u}}

\renewcommand{\d}{\mathrm{d}}

\newcommand{\koniec}{\begin{flushright}  $\Box $ \end{flushright}}
\def\be{\begin{equation}}
\def\ee{\end{equation}}

\def\u{\bf u}
\def\theequation{\thesection.\arabic{equation}}

\def\t{\tilde}
\def\wt{\widetilde}

\def\sm{\sigma}
\def\vv{\varepsilon}
\def\vp{\varphi}

\def\p{\partial}
\def\ov{\overline}

\def\lra{\longrightarrow}

\def\ra{\rightarrow}
\renewcommand{\d}{\mathrm{d}}
\def\h{\hat}
\def\b{\bar}
\def\om{\omega}
\def\mc{\mathcal}
\def\lmt{\longmapsto}
\def\dsl{\displaystyle}

\def\a{\alpha}
\def\ll{\lambda}

\newtheorem{theo}{Theorem}[section] 
\newtheorem{prop}[theo]{Proposition}  
\newtheorem{lemma}[theo]{Lemma}
\newtheorem{defi}[theo]{Definition}

\def\theequation{\thesection.\arabic{equation}}

\begin{document}
\title{
{\bf Compactified Twistor Fibration and Topology of Ward Unitons \vskip 20pt}}

\vskip 100pt

\author{
Prim Plansangkate\thanks{
  Email: prim.p@psu.ac.th} \\[15pt]
{\it Department of Mathematics and Statistics}, \\
 {\it Prince of Songkla University}, \\
{\it Hat Yai, Songkhla, 90110 Thailand}.}

\date{}
\maketitle

\vskip 50pt

\begin{abstract}
We use the compactified twistor correspondence for the $(2+1)$--dimensional integrable chiral model to prove a conjecture of
Ward.  In particular, we construct the 
 correspondence space of a compactified twistor fibration  and use it to 
 prove  that the second Chern numbers of the holomorphic vector bundles, corresponding to the 
uniton solutions of the integrable chiral model, equal the third homotopy classes of the restricted extended solutions of the unitons.
Therefore we deduce that the total energy of a time-dependent uniton is proportional to the second
Chern number.

\end{abstract}

\newpage
\setcounter{equation}{0}

\section{Introduction}

The integrable chiral model, also known as the Ward chiral model, was first introduced by Ward
in \cite{W88} 
as a rare example of an
integrable system in $(2+1)$ dimensions that
{admits} soliton solutions,  and yet is close to being Lorentz invariant.    
Many time-dependent soliton solutions - solutions which have smooth energy
density concentrated in some {finite} region of space  -
have been constructed explicitly \cite{W88, W95, I96, IZ98, DT04}.  Some  represent multi-soliton configurations 
 with no scattering, while some represent solitons which do scatter.  Analytic and algebraic B\"acklund transformations 
which can be used to generate all soliton solutions have also been developed \cite{IZ98, DT04}. 

The integrability of the Ward chiral model  comes from the fact that the equation can be realised 
as  a symmetry reduction of the
 anti-self-dual Yang-Mills (ASDYM) equation in $(2+2)$ dimensions.  Hence, it is integrable by twistor method   
(see, for example, \cite{WW, MW, MDbook}).
  Ward gave a one-to-one correspondence between solutions of the Ward chiral model to certain 
holomorphic vector bundles over the minitwistor space \cite{W89, W90}.  The minitwistor space is a 2-dimensional complex manifold.  
It is in fact the holomorphic tangent bundle, $T\bP^1,$ of the Riemann sphere $\CP^1.$

This paper deals with a particular class of solutions of the Ward chiral model.  These solutions correspond to the holomorphic vector bundles which
extend to a fibrewise compactification $\ov{T\bP^1}$ of  $T\bP^1.$  \, In \cite{W90, W96} Ward discussed this class
of solutions and the boundary conditions they have to satisfy, in order for the corresponding vector bundles to extend
to $\ov{T\bP^1}$.  This correspondence was formulated as a theorem and proved by
Anand \cite{An, Anand98}.   The aim of this paper is to answer a question posed by Ward in \cite{W90}, of what topological invariant in spacetime
 corresponds to the second Chern number of the holomorphic vector bundle over the 
compact space $\ov{T\bP^1}.$  \, It was stated in both \cite{W96, Anand98} that the second Chern number 
 corresponds to a topological degree of the Ward chiral field, taking values in $\pi^3(U(N)),$  however no proof was presented.
The goal of  this paper is then to provide a proof of this conjecture, thus giving an interpretation of the second Chern number in spacetime.

\vspace{0.5cm}

The Ward chiral model is given by
\be \label{Wardeq}
(J^{-1}J_t)_t-(J^{-1}J_x)_x-(J^{-1}J_y)_y-[J^{-1}J_t,
J^{-1}J_y]=0, \ee
where $J:\R^{3}\longrightarrow U(N)$, \,
$(x, y, t)$ are coordinates on $\R^3$ with the line element \\$\eta =  dx^2 + dy^2 - dt^2,$ and $J_x:=\p_x J,$ etc.

The model is in fact 
equivalent to the Yang-Mills-Higgs (YMH) system  in $(2+1)$ dimensions, with a gauge fixing.  
Let   $A=A_t d t + A_x d x + A_y d y$ and   $\Phi$  be a one-form and a function
on $\R^{2,1}$ respectively, with values in the Lie algebra $\uu(N).$  The YMH system is given by
\vspace{-0.5cm}
\begin{eqnarray} \label{YMHeqns}
\p_x \Phi + [A_x, \Phi] &=& \p_y A_t - \p_t A_y + [A_y, A_t], \nonumber \\
\p_y \Phi + [A_y, \Phi] &=& \p_t A_x - \p_x A_t + [A_t, A_x],  \\
\p_t \Phi + [A_t, \Phi] &=& \p_y A_x - \p_x A_y + [A_y, A_x], \nonumber 
\end{eqnarray}
where the gauge potential $A$  and the
Higgs field $\Phi$ are 
 determined up
to gauge transformations
\[
A\longrightarrow b^{-1}A\,b + b^{-1} d b , \qquad
\Phi\longrightarrow b\,\Phi\, b^{-1}, \qquad b=b(x^\mu)\in U(N).
\]
The system (\ref{YMHeqns}) reduces to the Ward chiral model (\ref{Wardeq}) under a gauge choice 
\be \label{gaugechoice}  A_t=A_y=\frac{1}{2}J^{-1}(J_t+J_y), \qquad
A_x=-\Phi=\frac{1}{2}J^{-1}J_x, \ee
where $J$ is a $U(N)$-valued function.

\vspace{0.5cm}

Now, since the YMH system is a reduction of the ASDYM equation by a non-null translation, it inherits a 
twistor correspondence, established by Ward in \cite{W89}.   This gives a one-to-one correspondence between solutions
of the YMH system (\ref{YMHeqns}) and holomorphic vector bundles over the minitwistor space.  As mentioned earlier,
the minitwistor space is the holomophic tangent bundle $T\bP^1$ of the Riemann sphere.  It is considered to be the 
space of null planes in
$\C^3$ - a complexification of $\R^{2,1}$.   That is, a point of $T\bP^1$ corresponds to a null plane in $\C^3.$
On the other hand, a point in $\C^3,$ including points in $\R^{2,1}$ as  a real slice, corresponds a holomorphic section,
a $\CP^1$ line, of $T\bP^1.$  We shall refer to a holomorphic section $\h p$ associated with a point $p \in \R^{2,1}$ 
 as  a real section.

The twistor correspondence for the Ward chiral model follows
  readily from the correspondence for
the YMH system.  However, since the field $J$ is obtained from $(A,
\Phi)$ by integration, it contains more information than the YMH
fields.  This additional data is a holomorphic
framing of the vector bundle  along two fibres of $T\bP^1$  \cite{W90}.   

\vspace{0.5cm}

Ward also discussed in particular the static solutions of (\ref{Wardeq}).  These are harmonic maps from $\R^2$ to 
$U(N).$  The finite energy condition allows the harmonic maps to extend to $S^2.$  Moreover, it was shown, for the gauge group
$SU(2),$ that finite energy static solutions correspond to holomorphic vector bundles which extend to a fibrewise 
compactification $\ov{T\bP^1}$ of $T\bP^1.$  The generalisation of this to the gauge group $U(N)$ was established 
by Anand in \cite{An0}. 

A class of time-dependent solutions of (\ref{Wardeq}) was also shown in \cite{W90} to give rise to holomorphic vector bundles
over $\ov{T\bP^1}.$  To discuss this class of solutions, one needs to consider a Lax formulation of (\ref{Wardeq}).

\vspace{0.5cm}

The integrability of the Ward chiral model  can be realised from the
fact that it is the compatibility condition  for  a system of overdetermined linear equations, the so-called Lax pair.
The Lax pair for (\ref{Wardeq}) comes naturally from the Lax pair of the YMH system  (\ref{YMHeqns}):
\be
\label{laxpair}
L_0\Psi := (D_y+D_t-\ll(D_x+\Phi))\Psi=0,\qquad
L_1\Psi := (D_x-\Phi-\ll(D_t-D_y))\Psi=0,
\ee
where $D_x:= \p_x + A_x$ is the covariant derivative with respect to the gauge field $A_x,$ similarly for
$D_y$ and  $D_t,$ and   $\Psi$ is a $GL(N, \C)$-valued function of the spacetime coordinates $(x,y, t)$ and a complex
parameter $\ll\in \CP^1.$  The YMH system (\ref{YMHeqns}) arises as the compatibility condition $[L_0, L_1]=0.$

The gauge freedom of (\ref{YMHeqns}) allows us to choose 
the gauge (\ref{gaugechoice}), in which the Ward chiral model becomes the compatibility condition for the Lax pair (\ref{laxpair}).
That is, if the map $J$ in (\ref{gaugechoice})  is a solution of (\ref{Wardeq}), then the overdetermined  system 
(\ref{laxpair}) admits a matrix solution $\Psi(x, y, t, \ll)$ which satisfies the unitary reality condition
\be \label{unitrealcond}
\Psi(x, y, t, \ov{\ll})^* \, \Psi(x, y, t, {\ll})={\bf I},
\ee
where ${\bf I}$ is the identity matrix.
On the other hand, a solution $\Psi$ of (\ref{laxpair}) which satisfies (\ref{unitrealcond}) gives rise to a solution of (\ref{Wardeq}) via
\be \label{JfromP} J(x,y,t)=\Psi^{-1}(x,y,t, 0), \ee
and all solutions of (\ref{Wardeq}) can be constructed in this way.  (See, for
example, \cite{HSW99}.)  
 The matrix solution $\Psi$  is called
an extended solutions.

The sufficient conditions for the holomorphic vector bundles corresponding to Ward chiral fields to extend to 
 $\ov {T\bP^1},$ was first discussed by Ward in \cite{W96}.  In addition to the finite energy
condition, ensured by the boundary condition (valid for all $t$)
\be
\label{assympt}
J=J_0+J_1(\varphi)r^{-1}+O(r^{-2})\qquad
\mbox{as}\qquad r\longrightarrow \infty,\qquad x+iy=re^{i\varphi},
\ee
one also needs a global boundary condition on the extended solution $\Psi$ of the Lax pair (\ref{laxpair}). 
Recall that $\Psi(x, y, t, \ll)$ is defined on $\R^{2,1} \times \CP^1.$ \, Let $\psi$ be the restriction of the map
$\Psi$ to the spacelike $t=0$ plane and the real equator $S^1 \subset \CP^1$ of the space of spectral parameter $\ll,$ i.e. 
\be \label{restext}
\psi(x, y, \theta) := \Psi \left(x, y, 0,-\cot \left(\frac{\theta}{2}\right)\right), 
\ee
where we have made change of variable for real $\ll = -\cot \left(\frac{\theta}{2}\right).$ Then the global
boundary condition, the so-called ``trivial scattering" boundary condition\footnote{The restricted extended solution $\psi$ satisfies 
$ \displaystyle \label{holonomy2} (u^\mu D_\mu-\Phi)\psi=0,$
where the operator anihilating $\psi$
is the spatial part of the Lax pair (\ref{laxpair}), given by
\[
\label{holonomy} \frac{\ll L_0+ L_1}{1+\ll^2}=u^\mu D_\mu-\Phi,
\qquad \mbox{where}\qquad {\u}=\Big(0, \frac{1-\ll^2}{1+\ll^2},
\frac{2\ll}{1+\ll^2}\Big) = (0, -\cos{\theta}, -\sin{\theta}).
\]
The trivial scattering condition (\ref{trivial_scatt}) implies that the differential operator
$u^\mu D_\mu-\Phi$ has trivial monodromy  along the compactification $S^1$ of
a straight line $ (x,y) = (x_0-\sigma \cos{\theta},y_0-\sigma \sin{\theta}), \sigma\in\R$.},
is given by
\be \label{trivial_scatt} \psi (x,y, \theta)
\longrightarrow \psi_0(\theta) \qquad \mbox{as} \qquad r = \sqrt{x^2+y^2}
\longrightarrow \infty, 
\ee 
where $\psi_0(\theta)$ is a  
$U(N)$-valued function on $S^1$.

It was described in \cite{W96}, particularly for the gauge group $SU(2),$ that the class of Ward
chiral fields, for which the corresponding vector bundles $E$ extend to  $\ov{T\bP^1},$
are those which satisfy the boundary conditions (\ref{assympt}) and (\ref{trivial_scatt}).  This was then 
formulated as a correspondence by Anand for a general gauge group $U(N).$
\, The compactified
minitwistor space $\ov {T\bP^1}$ was defined in \cite{W90}
to be the fibrewise compactification of
$T\bP^1$  where each $\C$-fibre becomes a copy of $\CP^1.$  One can
also think of $\ov {T\bP^1}$ as a cone in $\CP^3$ with blown-up
vertex. 

\vspace{0.5cm}

\noindent {\bf Ward-Anand correspondence}   {\cite{W96, An}} \; 
There  is a one-to-one correspondence
between
\begin{enumerate}[(i)]
\item Real-analytic solutions $J: \R^{2,1} \ra U(N)$ of the Ward chiral model (\ref{Wardeq}) which satisfy the boundary conditions
(\ref{assympt}) and (\ref{trivial_scatt}),
and 
\item Holomorphic rank-$N$ vector bundles $E$ over the
compactified minitwistor space $\ov{T\bP^1},$ such that $E$ satisfies certain reality conditions
and when restricted to real sections and to the fibres of $\ov{T\bP^1}$ over the real equator $S^1 \subset \CP^1$
of the base, $E$ is trivial with a fixed framing.
\end{enumerate}

\vspace{0.5cm}

The  holomorphic vector
bundle ${E \ra \ov {T\bP^1},}$ which now fibres over a compact manifold,  has Chern numbers as 
topological invariants.  The fact that the bundle is trivial when restricted to real sections  implies that the first Chern
number vanishes.  The next non-trivial invariant is the second Chern number.  On the other hand, the finite energy Ward chiral
fields which satisfy the trivial scattering condition 
admit a well defined topological
degree, associated with their extended solutions.  
The boundary conditions (\ref{assympt}) and (\ref{trivial_scatt}) enables $\psi$ to
 extend to the suspension $SS^2 = S^3$ of $S^2.$  (See \cite{DP06}.)  The restricted extended solutions $\psi,$ now as 
 maps from $S^3$ to $U(N),$ are classified by the third homotopy class \cite{topol} \be
\label{degree1}
[\psi]=\frac{1}{24\pi^2}\int_{S^3}\mbox{Tr}((\psi^{-1}\d \psi)^3),
\ee 
which is an integer taking values in
$\pi_3(U(N))=\Z$ and invariant under continuous deformations
of $\psi$.   \\

The identification between the topological degree of 
the extended solutions  and the second Chern number of
the corresponding vector bundles was stated in both \cite{W96} and \cite{Anand98},
however the proof was not presented.  This conjecture is supported by the works
in \cite{An0, DP06}.  In \cite{An0}, Anand showed that the energy of a static solution of the $U(N)$ Ward chiral model 
is proportional to the second Chern number of the corresponding vector bundle.  Later in \cite{DP06}, a class of Ward chiral fields called 
time-dependent unitons, which includes the static solutions, was shown to have their total energy proportional to the third homotopy class
of the extended solutions.  This explains an observation of discrete total energy of time-dependent unitons in \cite{IM04}.
 It also gives the identification between the third homotopy class
of the extended solutions and the second Chern number of the holomorphic vector bundles for finite energy static Ward chiral fields.  \\

The time-dependent unitons form a class of solutions which satisfy the boundary conditions (\ref{assympt}) and (\ref{trivial_scatt}). 
 These are soliton solutions for which  the  extended solutions 
have a pole of arbitrary order 
multiplicity in the complex plane of the spectral parameter $\ll.$
In \cite{DT04}, Dai and Terng
 have demonstrated that an extended solution
satisfying the trivial scattering condition has
poles at non-real points $\mu_1, ..., \mu_r,$ with multiplicities
$n_1, ..., n_r$, and is a product of  $r$ $N \times N$ matrices, called simple elements.  The identification of the two topological invariant is
valid for any such solutions.  However, only the time-dependent unitons, where $r=1,$ have their energy directly proportional to the third
homotopy class $[\psi].$ \\

The main result in this paper is the following theorem.

\begin{theo}
\label{main_th}
Let $\ov {T\bP^1}$ be the fibrewise compactification of $T\bP^1$
where each fibre becomes $\CP^1,$ and 
${E \ra \ov {T\bP^1}}$ be the  holomorphic vector
bundle, corresponding to a solution of the $U(N)$ Ward chiral model {\rm(\ref{Wardeq})} which satisfies the boundary conditions
{\rm(\ref{assympt}), (\ref{trivial_scatt})}.   Let $c_2(E)$ be
the second Chern number of  $E,$ given by 
\[ c_2(E) = -\frac{1}{8 \pi^2}  \int_{\ov{T\bP^1}} \mbox{Tr} (F \wedge F), \]
where $F$ is the curvature two-form of an arbitrary connection on $E,$ and $[\psi]$ be the third homotopy class of the restricted extended solution,
defined in {\rm (\ref{degree1})}.  

Then
\[  c_2(E) = [\psi]. \]
\end{theo} 

\vspace{0.5cm}

Our proof of Theorem \ref{main_th} is based on the existence of a double fibration from
 a space we shall call the restricted correspondence
space ${\mc F},$ to a compactification of a real spacelike plane ($t=0$) $\R^2 \subset \R^{2,1}$ 
(where the restricted extended solution $\psi$ is defined) and the  compactified minitwistor space    $\ov {T\bP^1}.$
Consequently, the vector bundle ${E \ra \ov {T\bP^1}}$ can be pulled back to a bundle $E^* \ra {\mc F},$
for which the second Chern number can be calculated and related to the topological degree of $\psi.$  
\, A related problem has been considered by Mason in \cite{Mason04}, where an initial value problem for
the Ward chiral model was formulated on a null hypersurface.

The structure of the paper is as follows.
In Section 2 we  give a detailed exposition 
 of the twistor correspondence in the holomorphic setting, which was described briefly in
\cite{W96}, between  the compactified spacetime $\ov{M_\C} =
 \CP^3$ and $\ov {T\bP^1}.$ \, Starting from the identification between the
minitwistor space $T\bP^1$ and a cone ${\mc C}$ minus its vertex in another complex
projective 3-space $\CP^{3*}$, we explain how the 
compactification $\ov {T\bP^1}$ of $T\bP^1$ is equivalent to the cone ${\mc C}$ with blown-up
vertex.  

Then in Section 3, we construct the restricted correspondence space
 for a double fibration over $\ov{M_\C}$ and $\ov {T\bP^1}$.  A double fibration picture was discussed in
\cite{An0},  where the correspondence space was taken
to be a
singular variety in the direct product $\ov{M_\C} \times \CP^{3*}$  and one of the
target spaces is the cone ${\mc C}$ instead of $\ov {T\bP^1}.$ 
\, Here we explore a double 
fibration where the correspondence space is a blow up of the
singular variety, which fibres over $\ov{M_\C}$
and  $\ov {T\bP^1}.$  Then we define the 
restricted correspondence space which fibres over an $\RP^2,$ regarded
as a compactification of a spacelike plane $\R^2 \subset \R^{2,1},$ and
show that it admits  a surjective map to $\ov {T\bP^1}.$  
Finally we give a proof of Theorem \ref{main_th} in Section 4.


\section{Compactified minitwistor space}  \label{sec:compminitwist}

The twistor geometry for $(2+1)$ dimensional flat spacetime and its use to construct the YMH fields in $\R^{2,1}$
was introduced in \cite{W89}.  The minitwistor space can be thought of as the space of null planes in $\C^3,$
considered as a complexification of $\R^{2,1}.$  
Let $(x,y,t)$ be complex coordinates on the complexified spacetime $M_\C = \C^3.$ 
A null plane in $M_\C$ is given by 
\be \label{section} \om = 2x\ll + y(\ll^2 -1) + t(1 +\ll^2),  
\ee
where $\om \in \C$ and $\ll \in \CP^1$ are complex parameters.

Thus the space of null planes is a 2-dimensional complex manifold, which is actually the 
holomorphic tangent bundle of the Riemann sphere $\CP^1.$  Hence, we denote the 
minitwistor space by $T\bP^1,$ with $\om$ and $\ll$ as fibre and base coordinates, respectively.  
See  \ref{appa} for details.  (The minitwistor space $T\bP^1$ was used to construct static YMH monopoles
on Euclidean space $\R^3 \, \cite{Hitchin82}.$)

Another picture of the 
  minitwistor space was given in \cite{W90, W96} as a cone minus its vertex in a complex 3-dimensional 
projective space.  The cone picture proves to be convenient
in the study of the compactified double fibrations, which will be essential to our proof of Theorem \ref{main_th}.
  Therefore in this section, we
shall start with a detailed explanation of how to identify $T\bP^1$ with 
 a cone ${\mc C},$
 without the vertex.  Then we shall discuss the correspondence between points in spacetime $M_\C$ and points on the cone
${\mc C}.$  This will lead to a natural compactification of spacetime, $\ov{M_\C} = \CP^3,$ and 
the identification of the blow up of the cone ${\mc C}$ with the compactified minitwistor space $\ov{T\bP^1}.$\\

Let us first define the cone ${\mc C}$.   Let $Z_\a =
(Z_0, Z_1, Z_2, Z_3 ) \in \C^4-\{0\}$ be homogeneous
coordinates of a complex projective 3-space, denoted by $\CP^{3*}.$  
  A cone  ${\mc C}$ in $\CP^{3*}$ is given by
\be \label{homocone} ({Z_1})^2 + ({Z_2})^2 - ({Z_3})^2 = 0. \ee
Note our convention of one minus sign. The vertex is the point ${\bf
  z_0} = [Z_0,0,0,0], \; Z_0 \ne 0.$
To see that there is a bijection between ${\mc C} - \{{\bf z_0}\}$ and $T\bP^1,$ note that 
for any point on the cone except the vertex, one can parametrise
  $Z_i := (Z_1, Z_2, Z_3) \ne (0,0,0)$ by  $\pi^A  \in \C^2 - \{0\}$ as follows.  Let
\[
Z^{AB}=  \left (
\begin{array}{cc}
\frac{Z_3+ Z_2}{2} & \frac{Z_1}{2}\\
\frac{Z_1}{2} & \frac{Z_3 - Z_2}{2}
\end{array}
\right ).
\] 
Then equation (\ref{homocone}), which is the same as $-4\; \mbox{det}(Z^{AB}) = 0,$
implies that $\mbox{rank}\; {(Z^{AB})=1}.$  Hence
\[ Z^{AB} = \pi^A \pi^B = \left( 
\begin{array}{cc}
({\pi^0})^2 & \pi^0 \pi^1 \\
\pi^0 \pi^1 & ({\pi^1})^2
\end{array}     
\right),
\]   
where $\pi^A  \in \C^2 - \{0\}.$
In other words, one can parametrise solutions of (\ref{homocone}) 
 with $Z_i \ne (0,0,0)$ by  
\be \label{homoparamcone}
Z_{\a} =(\hat \om, \;  -2\pi_0 \pi_1,\; {\pi_1}^2 - {\pi_0}^2, \; 
{\pi_0}^2+{\pi_1}^2 ),
\ee
where  $\hat \om \in \C$
 is arbitrary, $\pi_A = \pi^B \vv_{BA}$ and $\vv_{BA}$ is the
 alternating symbol.   This shows that points $[Z_\a] \in {\mc C} -
\{{\bf z_0}\}$ can be parametrised inhomogeneously in two patches, where $\pi_1 \ne 0$
and where $\pi_0 \ne 0,$ by 
\be \label{paramcone}  [\om, -2\ll, 1-\ll^2, \ll^2+1 ] \quad \mbox{and} \quad 
 [\t\om, -2\t\ll, \t\ll^2-1, 1+\t\ll^2 ],
\ee
respectively, where $\dsl \om :=
\frac{\hat \om}{\pi_1^2}, \; \ll := \frac{\pi_0}{\pi_1}$  and $\dsl \t \om :=
\frac{\hat \om}{\pi_0^2}, \; \t \ll := \frac{\pi_1}{\pi_0}.$
In the overlap, the inhomogeneous coordinates are related by $\dsl \tilde \ll
= \frac{1}{\ll}$ and $\dsl \tilde \om = \frac{\om}{\ll^2}.$ \,  This gives the
equivalence between ${\mc C} - \{{\bf z_0}\}$ and $ {T\bP^1}.$  
That is, there exists a bijection from  $T\bP^1$ to ${\mc C} - \{{\bf z_0}\}$ given locally in the
two patches by 
\be \label{map1}
(\om, \ll) \mapsto [\om, -2\ll, 1-\ll^2, 1+\ll^2 ] \quad   \mbox{and} \quad
(\tilde \om, \tilde \ll) \mapsto [\tilde \om, -2\t \ll, {\t \ll}^2-1,
  1+{\t \ll}^2 ].
\ee
  In fact, (\ref{map1}) gives a biholomorphism
 between $T\bP^1$ and ${\mc C} - \{{\bf z_0}\} \subset \CP^{3*}.$  \\

Let us now describe the correspondence between the complexified
spacetime ${M_\C = \C^3}$ and the minitwistor space in the cone picture.
For convenience, let us think of $M_\C$ as embedded in a
$\CP^3.$   Let $P^{\a} = (P^0, P^1, P^2, P^3) \in \C^4-\{0\}$ be homogeneous
coordinates on the $\CP^3$ and take the open set $P^0 \ne 0$ to be our
spacetime $M_\C.$  A plane in $\CP^3$ is defined to be the projection
of a 3-dimensional subspace of the associated $\C^4,$ given by 
\be \label{homoplane}
Z_0P^0 +Z_1P^1+Z_2P^2 - Z_3P^3 = 0.
\ee
Note again our convention of one minus sign.  Each plane is thus labelled by ${Z_{\a} \in \C^4-\{0\}}$ up to a
constant multiplication.  That is, the space of planes in $\CP^3$ is
another complex projective 3-space, which we shall denote $\CP^{3*}.$
   Then in this setting, 2-planes in $M_\C$ are the $\C^2$-intersections
of planes in $\CP^3$ with $M_\C.$ 
 
This picture suggests a natural compactification of the spacetime $M_\C$
to $\ov {M_\C} = \CP^3.$  One can think of $\ov {M_\C}$
as $M_\C + \CP^2,$ where $\CP^2$ is the complement region $P^0
=0.$  \, Let  
\be \label{xytcoord} x=\frac{P_1}{P_0}, \quad y=\frac{P_2}{P_0}, \quad
 t=\frac{P_3}{P_0} \ee 
be coordinates on $M_\C.$  Then, one can interpret the
complement $\CP^2$ as the infinity boundary, which will be denoted by
$\CP^2_\infty.$  To make contact with a real setting, note that since  $\CP^2
\cong S^5 / S^1,$ the $\CP^2_\infty$ can then be thought of as the $S^5$ infinity boundary of
$M_\C \cong \R^6$ with the points on $S^1$ orbits identified.  \\

\begin{defi}  
 A plane {\em (\ref{homoplane})} in $\ov {M_\C}
= \CP^3$ is called a null plane if $[Z_\a] \in \CP^{3*}$ lies in
the cone ${\mc C}$ {\em (\ref{homocone})}. 
\end{defi}

Let us now show that null planes in $\ov {M_\C}$  give rise to null
planes in $M_\C,$ as defined in \ref{appa}.  Since in $M_\C,$ $P^0
\ne 0,$  one can divide (\ref{homoplane}) by $P^0$ and use the
coordinates (\ref{xytcoord}).  By substituting in the first 
parametrisation of (\ref{paramcone}) for $Z_\a,$ equation (\ref{homoplane}) becomes the null plane equation 
(\ref{section})
\[ \om = 2x\ll + y(\ll^2 -1) + t(1 +\ll^2).  
\]
Note that the
parametrisation (\ref{paramcone}) is only valid for the points on ${\mc C} -
\{{\bf z_0}\}.$  From the plane equation (\ref{homoplane}), one sees
that  the vertex ${\bf z_0} = [Z_0, 0,0,0]$ corresponds to the
infinity boundary $\CP^2_\infty,$ which we shall regard as a null
plane by definition.   Hence, the natural extension from $M_\C$ to
$\ov{M_\C}$ makes the inclusion of the vertex ${\bf z_0}.$   

\vspace{0.5cm}

The correspondence between points in the compactified spacetime $\ov {M_\C}
= \CP^3$ and points on the cone ${\mc C} \subset \CP^{3*},$ including the vertex ${\bf z_0},$ is summarised in
Lemma \ref{cp3andconelem} below.  First, let us define what we mean by
a conic section of ${\mc C}.$

\begin{defi}
A conic section of a cone ${\mc C} \subset \CP^{3*}$  is given by the
intersection of a plane in $\CP^{3*}$ with ${\mc C}.$
\end{defi}

\vspace{0.1cm}

\begin{lemma}
\label{cp3andconelem}
There is a one-to-one correspondence between points on 
the cone minus the vertex,  ${\mc C} - \{{\bf z_0}\} \subset \CP^{3*},$
and  null planes in $M_\C = \C^3 \subset \ov {M_\C} = \CP^3.$ 
The vertex ${\bf z_0}$ corresponds to the infinity boundary
$\CP^2_\infty \subset \ov {M_\C}.$  

\vspace{0.5cm}

On the other hand, there is a one-to-one correspondence between points
in $\ov {M_\C}$ and conic sections of ${\mc C},$ where
\begin{enumerate}[{\rm 1.}]
\item points in $M_\C$ correspond to the conic sections that do not
intersect ${\bf z_0}$ 
\item points in $\CP^2_\infty$ correspond to the conic sections, each
of which consists of two $\C$-lines, counting multiplicity, meeting at ${\bf z_0}.$
\end{enumerate}
\end{lemma}

\vspace{0.5cm}

\noindent {\bf Proof.} \; We have already established the first part of
the lemma.  The second part can be proved by considering equation 
(\ref{homoplane}).  By fixing
  $[P^\a]$ and varying $[Z_\a]$ instead, one sees that (\ref{homoplane}) is also the
  equation for planes in $\CP^{3*}.$  That is, a point $[P^\a] \in
 \ov{M_\C} = \CP^3$ labels a plane in $\CP^{3*}.$  Moreover, for a given $[P^\a]$ it is
always possible to find common solutions
$[Z_\a]$ to (\ref{homocone}) and (\ref{homoplane}), which means that any plane in
$\CP^{3*}$ intersects ${\mc C}.$  

Now, since $P^0
  \ne 0$ for a point in $M_\C,$ no plane labelled by $[P^\a] \in M_\C$ passes through
  ${\bf z_0}.$  Hence we have that each point in $M_\C$
  corresponds to a conic section on ${\mc C} - \{{\bf z_0}\}.$  
 For a point on
$\CP^2_\infty,$ with $P^0 = 0$ the corresponding plane in $\CP^{3*}$ is given by
\be \label{inftysection} P^1Z_1+P^2Z_2 - P^3Z_3 = 0. \ee
Equation (\ref{inftysection}) admits the vertex $[Z_0, 0, 0, 0]$ as a
solution.  Thus, the plane passes through the vertex ${\bf z_0}.$  
Thinking analogously of a cone in
$\R^3,$ one would expect the conic section to consist of two lines coming
together at the vertex.  This is indeed the case.  
For $(Z_1, Z_2, Z_3) \ne (0, 0, 0),$ we can use the parametrisation
(\ref{homoparamcone}) to label $Z_i.$  For concreteness, let us
consider the patch where $\pi_1 \ne 0$ and use the first
parametrisation in (\ref{paramcone}).  Equation (\ref{inftysection}) becomes
\be \label{tworoots} (P^2 +P^3) \ll^2 + 2P^1 \ll + (P^3-P^2)=0.  \ee
This is a quadratic equation for $\ll.$ \, Since $\om$ (corresponding to
$Z_0$) is arbitrary, it implies that a conic section corresponding to
a point on $\CP^2_\infty$
consists of two $\C$-lines of constant $\ll,$ whose values are given by the two roots of
(\ref{tworoots}) counting multiplicity.  In the limit where $\om$
approaches infinity, the two lines meet at ${\bf
  z_0}.$ \koniec

The cone ${\mc C}$ is however not equivalent to 
the compactified minitwistor space defined 
in \cite{W90}.  The compact space $\ov {T\bP^1}$ is defined to be the fibrewise compactification  
 of $T\bP^1,$ where each fibre is
extended from $\C$ to $\CP^1.$  This can be regarded as adding a $\CP^1$
at $\om = \infty.$  We shall denote the additional $\CP^1$ by
$L_\infty.$   In the cone picture, the compactified minitwistor space is
the cone ${\mc C}$ with the vertex blown up to a $\CP^1.$ 

\vspace{0.5cm}

\begin{prop}
\label{blowz0andLinfty}
There exists a bijection from
$\ov {T\bP^1}$  to the blow-up $\wt {\mc C}$ of the cone ${\mc C}
\subset \CP^{3*}$ at the vertex ${\bf z_0},$ where the blow-up of
${\bf z_0}$ is identified with $L_\infty.$
\end{prop}

\vspace{0.2cm}

A detailed calculation of the blow up and a proof of Proposition \ref{blowz0andLinfty}
can be found in \ref{appb}.  \\

Lastly, let us recall briefly the construction of holomorphic vector bundles over the minitwistor space
from solutions of the Ward chiral model.  This can be done in holomorphic setting where $\R^{2,1}$
is complexified to $M_\C =\C^3.$  Consider the Ward chiral model (\ref{Wardeq}) in $M_\C.$  Suppose $J$ in (\ref{gaugechoice})
satisfies (\ref{Wardeq}).  Then, for a fixed $\ll,$ there exists $N$ linearly independent column vector solutions of 
the Lax pair (\ref{laxpair}), forming the fundamental $N \times N$ matrix solution  $\Psi.$  These 
column vector solutions are covariantly constant sections of the trivial $\C^N$ bundle $V \ra M_\C$ 
restricted to null planes, with respect to  $(A_\mu, \Phi)$ in (\ref{laxpair}).
Let $z \in T\bP^1$ corresponds to a null plane $Z \subset M_\C.$
Then, one defines a holomorphic  rank $N$ vector bundle over $T\bP^1$ by taking the fibre over each point $z \in
T\bP^1$ to be the space of covariantly constant sections of $V|_Z.$

The Ward chiral fields in $\R^{2,1}$ then correspond to such holomorphic vector bundles with reality conditions.  This
is described in details in \cite{W90}, where patching matrices of $E \ra T\bP^1$ are given explicitly for 
static $SU(2)$ $1$-uniton solutions.  It was explained then how the patching matrices extend to $\ov{T\bP^1},$ thus defining
the bundle $E \ra \ov{T\bP^1}.$  This was generalised to the Ward-Anand correspondence \cite{An}.


\section{The Correspondence spaces}  \label{sec:corresspc}

The main aim of this paper is to prove Theorem \ref{main_th}, which identifies the second Chern
number of the holomorphic vector bundle $E$ over the minitwistor space $\ov {T\bP^1} $ and the third
homotopy class of the restricted extended solution $\psi$ to the Lax pair  (\ref{laxpair})
\[
(D_y+D_t-\ll(D_x+\Phi))\Psi=0,\qquad
(D_x-\Phi-\ll(D_t-D_y))\Psi=0,
\]
on $\R^{2,1}.$  

In the holomorphic setting,
 the extended solution $\Psi(x,y,t,\ll)$  is 
a function on the  correspondence space 
${\tt F} = \C^3 \times \CP^1$ of a double fibration over the spacetime $M_\C = \C^3$
and the minitwistor space $T\bP^1.$  Then $\psi$   is the restriction of $\Psi$ to certain real slice
$\R^2 \subset \R^{2,1} \subset \C^3$ and the real equator $S^1 \subset \CP^1$ of the space of 
spectral parameter $\ll.$  \, To relate the topological invariants on $\ov{T\bP^1}$ and that of $\psi,$ one could
think of pulling back the bundle over $\ov{T\bP^1}$ to ${\tt F}.$
However, as we cannot find a required
surjective map from ${\tt F}$ to $\ov{T\bP^1},$  we construct another
correspondence space, adapted to the compactified setting,
 which is the blown up version of that
presented in \cite{An0, An}.  Then we consider the restriction of  the correspondence space  to some
real and `$t=0$' slice.


\subsection{Compactified double fibration}

 Recall that the correspondence space in the non-compact
 double fibration  is the space of pairs of a spacetime point in $\C^3$ and a null
plane on which the point lies, where the null plane corresponds to a point on the
minitwistor space $T\bP^1.$  \, Hence ${\tt F}$ is a subset of $\C^3
\times T\bP^1$ defined locally by
\[ {\tt F} := \{(p,z) \in \C^3 \times T\bP^1 \; : \om = 2x\ll +
y(\ll^2-1) + t(1+\ll^2))  \},  \]
where $(x,y,t)$ are coordinates of a point  $p \in \C^3$ and $(\om,
\ll)$ are local coordinates of a point $z \in T\bP^1.$
 \, Given $(x, y, t) \in \C^3$ and $\ll
\in \CP^1,$ $\om$ is determined uniquely by the incidence relation,
and hence  ${\tt F}$ is biholomorphic  to $\C^3 \times \CP^1.$ 

\vspace{0.5cm}

For the compactified case, consider a singular algebraic
variety in $\CP^3 \times \CP^{3*}$ given by
\be \label{algebraic} 
\hat f := \{ (p,z) \in \CP^3 \times \CP^{3*} \; : Z_1^2+Z_2^2-Z_3^2
=0, \; P^0Z_0+P^1Z_1 + P^2Z_2-P^3Z_3 =0 \}.
\ee
This is effectively a subset of $\CP^3 \times {\mc C}$ which consists of
pairs of a point $p \in \CP^3$ and a point $z \in {\mc C}$ corresponding to
a null plane passing through $p.$    Equivalently, it is the set of pairs
of a point $z \in {\mc C}$ and a point $p \in \CP^3$ that corresponds to
a plane in $\CP^{3*}$  passing through $z.$  Hence, $\hat f$ has a 
natural double fibration  
\be \label{projectf} \hat r: \hat f \lra \ov {M_\C} = \CP^3 \qquad  \mbox{and} \qquad
\hat q:\hat f \lra {\mc C} \subset {\CP^{3*}}, \ee
 where $ \hat q \circ \hat r^{-1}(p)$ is the
 conic section $l_p \subset {\mc C}$ and $\hat r \circ\hat q^{-1}(z)$ is the null plane in
$\ov {M_\C}$ which is a $\CP^2,$ and ${\bf z_0}$ corresponds to $\CP^2_\infty.$  This is the
double fibration discussed in \cite{An0, An}.   

In the double fibration (\ref{projectf}), every point $z \in {\mc C}$
is on an equal footing: each point corresponds to a $\CP^2$ plane, including
${\bf z_0}.$  This is
not the case if one were to consider a similar fibration 
to $\ov {T\bP^1}.$

\vspace{0.5cm}

\begin{lemma} \label{Linftytocp1}
A point on $L_\infty \subset \ov {T\bP^1} \cong \wt {\mc C}$  corresponds to a $\CP^1
\subset \CP^2_\infty \subset \ov {M_\C}.$
\end{lemma}

\vspace{0.5cm}

\noindent {\bf Proof.} \;
First, note that since 
the finite points on $M_\C$ are holomorphic
sections (\ref{section}) in $T\bP^1 \subset \ov {T\bP^1}$ which do not
intersect $L_\infty,$ a point on $L_\infty$ must  
correspond to a subset of $\CP^2_\infty.$  This subset is determined by equation 
(\ref{tworoots}) for a fixed $\ll.$
Given a value of 
$\ll,$ (\ref{tworoots}) is one linear equation for 3 unknowns, $P^1, P^2, P^3.$
Since it is not possible for all coefficients to vanish at the
same time, one can always determine one variable in terms of the other
two. Hence, there are two degrees of freedom in the homogeneous
coordinates in $\C^2-\{0\},$ and we
conclude that each point in $L_\infty$ corresponds to a $\CP^1$ line
in $\CP^2_\infty.$   
 \koniec

\vspace{0.5cm}

We shall now present a double fibration which fibres over the
compactified minitwistor space $\ov {T\bP^1}$, where each point of $\ov {T\bP^1}$ has an equal
footing, i.e. a point on $L_\infty$ is also a $\CP^2.$  This is
achieved simply by defining the correspondence
space to be the blow-up of $\hat f$ along its singularity.  
The singularity of $\hat f$ comes from the conic
singularity ${\bf z_0} \in {\mc C},$ which  
corresponds to 
$\CP^2_\infty \times \{ {\bf z_0} \} \subset \h f.$  \\

That is, we define the correspondence space  $\hat {\mc F}$ of a double fibration
to the compactified spacetime $\ov {M_\C} = \CP^3$ and the
compactified twistor space $\ov {T\bP^1} \cong \wt {\mc C}$ as 

\begin{center} $\hat {\mc F} =$  {\it the blow-up of the algebraic variety $\hat f $
    {\em (\ref{algebraic})} along $\CP^2_\infty \times \{ {\bf
  z_0} \}.$}   \end{center}
The details of the blow up are given in \ref{appc}. \\

There exists a 
projection $\rho: \hat {\mc F} \lra \hat f$ such that, away from 
$\CP^2_\infty \times \{ {\bf z_0} \},$ $\rho$ is a one-to-one and onto.  
We find that the preimage of $\CP^2_\infty \times \{ {\bf z_0} \}$ under the map $\rho$ is
 $\CP^2_\infty \times L_\infty,$
where $L_\infty \in \wt{\mc C}$ is the blow up of ${\bf z_0}.$  \, Let us denote  
the preimage by $e:=\CP^2_\infty \times L_\infty.$ 

\vspace{0.3cm}

We define a surjective map $q: \hat {\mc F} \lra \wt{\mc C}$ by its action on two disjoint
regions. 
First, define 
\be \label{qdef1}  q|_{ \hat {\mc F}-e} : \; \hat {\mc F}-e \lra \wt{\mc C} - L_\infty\ee
such that  it is equivalent to the composition $\h q \circ \rho,$ where
$ \hat q: \hat
f - \CP^2_\infty \times \{ {\bf z_0} \} \lra {\mc C} - \{ {\bf z_0}
\} $
is the fibration in (\ref{projectf}).  \, Then, define
\be \label{qdef2} q|_{e} : \CP^2_\infty \times L_\infty \lra L_\infty\ee
to be the right projection. 
 Thus, by definition, $q$ is onto.  

\vspace{0.3cm}

Therefore, we have a double fibration 
\[ r: \hat {\mc F} \lra \ov{M_\C} = \CP^3 \quad  \mbox{and} \quad q:\hat {\mc F} \lra \wt {\mc C} = \ov {T\bP^1},\]
  where a finite spacetime point $p \in M_\C \subset \ov{M_\C}$
corresponds to a holomorphic section in $T\bP^1 \subset \ov {T\bP^1}$ and
a point $p \in \CP^2_\infty$ corresponds to the union of $L_\infty$ and two $\C$ lines of constant $\ll$ (counting multiplicity).
On the other hand, a point $z \in T\bP^1 \subset \ov {T\bP^1}$ corresponds to a null plane in $\ov{M_\C}$
(extension of a null plane in $M_\C$) and $z \in L_\infty$ corresponds to $ \CP^2_\infty.$


\vspace{0.5cm}

\subsection{The restricted correspondence space} \label{sec:restcorres}

Recall that the topological degree of a Ward chiral
field $J,$ satisfying the trivial scattering condition, comes from 
the third homotopy class of the restricted extended
solution $\psi(x,y, \theta).$  With this in mind we shall now define a
`restricted' correspondence space such that it gives rise to the 
domain of $\psi.$ 

Consider a `constant time' slice
$\tau,$  which is the $\CP^2 \subset \ov {M_\C}$ obtained by setting $P^3 =0.$ \, It is clear
that the intersection of $\tau$ with the noncompact spacetime
$M_\C = \C^3,$ where $P^0 \ne 0,$ is the $t=0$ \, $\C^2$-plane in the
$(x, y, t)$ coordinates (\ref{xytcoord}).  We will also consider
the `real slice' $\RP^3 \subset \ov{M_\C},$ which consists of the points
$[P^\a]$ whose homogeneous representatives can be chosen to be in $\R^4-\{0\}.$ Since
we write the line element on $M_\C$ as $d s^2 = d x^2 + d y^2 - d t^2,$ the
finite part of this $\RP^3$ is an $\R^{2,1}.$  Then, the $\RP^2$ intersection, denoted by 
$\tau_{\R},$ of $\tau$ with this $\RP^3$ can be thought of as
the extension of the $t=0$ $\R^2$-plane to the compactified space.  \\

We define the {\it restricted correspondence space} ${\mc F}$ to be the
restriction of $\hat {\mc F}$ to $\tau_\R.$  \,
This means that away from the singularity, ${\mc F}$ is the same as the
algebraic variety
\be 
\label{f3} 
f:=  \{ (p,z) \in \RP^3 \times \CP^{3*}: Z_1^2+Z_2^2-Z_3^2
=0, P^0Z_0+P^1Z_1 + P^2Z_2-P^3Z_3 =0, P^3 =0 \}
\ee
minus its singularity.  The singularity
of $f$ consists of the points $\{ ([P^\a], {\bf z_0}) \}$ for all $[P^\a]$ such that $P^0
=0.$ Since $P^3 =0,$ this is an $\RP^1 \subset \CP^2_\infty,$ which will be denoted by $\RP^1_\infty.$  
Under the usual projection map  $\rho: {\mc F} \ra f,$  the preimage  
of the singularity $\RP^1_\infty \times \{{\bf z_0}\}$ 
is $e_{\tau_\R} :=\RP^1_\infty \times L_\infty.$ \\

It is not immediate that the map $q$ defined by (\ref{qdef1}), (\ref{qdef2}) is still onto $\ov{T\bP^1} \cong \wt{\mc C}$ when 
the domain of $q$ is restricted to ${\mc F}$.  However, this turns out to be the case. \\

\begin{prop} \label{themapprop}
The restriction of the map $q: \h {\mc F} \lra \wt {\mc C}$ to ${\mc F},$
\be \label{themap} q|_{{\mc F}} : {\mc F} \lra \wt{\mc C},\ee
is surjective.
\end{prop}

\vspace{0.5cm}

\noindent {\bf Proof.} \,
First, it follows readily from (\ref{qdef2}) that 
\[ q|_{e_{\tau_\R}} : \RP^1_\infty \times L_\infty  \lra
L_\infty\]
is onto as a right projection.  However, it is not obvious that  
\be \label{qF-etr} q|_{{\mc F}-e_{\tau_\R}}:  {\mc F}-e_{\tau_\R} \lra \wt {\mc C} -
L_\infty \ee 
is also surjective.  The map (\ref{qF-etr}) is equivalent to the restriction of the map $\h q$ in (\ref{projectf}) 
to $f - \RP^1_\infty \times \{ {\bf z_0} \},$
\be \label{qf} \hat q:
f - \RP^1_\infty \times \{ {\bf z_0} \} \lra {\mc C} - \{ {\bf z_0}
\}. \ee
Therefore, the question whether $q|_{{\mc F}-e_{\tau_\R}}$ is onto comes
down to whether, given a point $[Z_\a] \in {\mc C} - \{{\bf  z_0}\},$ one can find a point in $\tau_\R$
which lies on the corresponding null plane.  In other words, whether all
null planes in $\ov {M_\C} = \CP^3$ intersect $\tau_\R.$  If this is the case, then we
can always find a (non-empty) preimage of every point in $ {\mc C} - \{{\bf  z_0}\}$ under (\ref{qf}), which
implies that (\ref{qf}) is onto, and hence so are (\ref{qF-etr}) and
(\ref{themap}).  

\vspace{0.3cm}

The intersection of a null plane in $\CP^3$ with $\tau_{\R}$ consists of points
  ${[P^\a] \in \RP^3}$ with $P^3=0$ satisfying
\be \label{intersect} 
Z_0P^0 + Z_1P^1 + Z_2P^2 = 0,
\ee
where $[Z^\a] \in \CP^{3*}$ labels the null plane, i.e. satisfies (\ref{homocone}). 
We proceed by direct calculation.  By looking for $[P^\a] \in
\tau_\R$ that satisfy (\ref{intersect}), we find that such solutions
exist for all $[Z_\a] \in {\mc C} - \{ {\bf z_0} \}.$  Therefore we conclude that the map
(\ref{themap}) is surjective.
The detailed calculation can be found in \ref{appd}.  
\koniec

\vspace{1cm}

What is crucial to our proof of Theorem \ref{main_th} is that now we have a
surjective map (\ref{themap}), which can be used to pull back the holomorphic vector  
bundle over $\wt{\mc C}$ to ${\mc  F}.$ 

We note here that the map $q|_{\mc F}$ is not one-to-one everywhere on ${\mc F}.$  
The preimage of a point $z\in L_\infty \subset \wt {\mc C}$ is the $\RP^1_\infty \times \{z\}.$  Then   
$\wt{\mc C} - L_\infty \cong {\mc C} - \{{\bf z_0}\}$ can be divided into two disjoint regions.  
Let ${\mc C}_\R$ denote the set of points $z :=[Z_\a] \in {\mc C}- \{{\bf z_0}\}$ whose representatives
can be chosen to be in $\R^4-\{0\}.$  We shall call the planes corresponding to $z \in {\mc C}_\R$ {\it real null planes}.
 Note that a real null plane with $(Z_1, Z_2, Z_3) \ne (0, 0, 0)$ corresponds to a null plane in $\R^{2,1}.$
It can be shown that the preimage in ${\mc F}$ of a point $z \in {\mc C}_\R$ is an $\RP^1.$  
Finally, the map $q|_{\mc F}$ is one-to-one
and onto ${\mc C}- \{{\bf z_0}\} - {\mc C}_\R.$  See \ref{appd} for details. \\

For the purpose of proving Theorem \ref{main_th} in the next section, we shall spend the final part of this
section describing ${\mc F}$ in $3$ disjoint regions:  First,

\vspace{0.3cm}

$\bullet$  \; $e_{\tau_\R}= \RP^1_\infty \times L_\infty.$  

\vspace{0.3cm}

 Then we divide
the complement ${\mc F} - e_{\tau_\R}$, which is identified with $f - \RP^1_\infty \times \{
{\bf z_0} \},$ into two regions: 

\vspace{0.3cm}

$\bullet$  \;  ${\mc R}= \{ (p,z) \in f : \, P^0 \ne 0, \, z \ne {\bf z_0} \}:$  This is the finite part where $p \in \R^2 \subset \tau_\R.$  
In fact since $P^0 \ne 0,$ we have that  $(Z_1, Z_2, Z_3) \ne (0,0,0).$  Hence,  
   $(Z_1, Z_2, Z_3) \in {\mc R}$ can be
parametrised by $[\pi_A] \in \CP^1.$    Now,  since
$P^0 \ne 0,$ one can use (\ref{intersect}) to determine $Z_0.$
That is, 
${\mc R}$ can be parametrised by $(P^1,P^2) \in \R^2 \subset \RP^2$ and
$[\pi_A] \in \CP^1.$  It can thus be deduced that ${\mc R} = \R^2 \times
\CP^1.$ This indicates, as we expect, that there are $\CP^1$-worth of null planes passing through each point $p.$

\vspace{0.3cm}

$\bullet$ \; ${\mc R_\infty}= \{ (p,z) \in f : \,  P^0 = 0, \, z \ne {\bf z_0}\}:$  
This region corresponds the extension of null planes
to $\RP^1_\infty.$

\vspace{1cm}

\noindent {\bf Remark.} The constant time slice $\tau$ and the
real slice $\RP^3$ used to define $\tau_\R$
can be regarded
 as the sets of fixed points of a holomorphic involution and an
 anti-holomorphic involution, respectively, on
the complexified spacetime $\ov{M_\C}.$ \, The maps induce the
corresponding involutions on the minitwistor space $\ov{T\bP^1}.$  
\, For details, see \ref{appe}.

\vspace{1cm}

\section{Chern numbers and topological degrees}

For a Ward chiral field $J$ satisfying the boundary conditions (\ref{assympt}) and (\ref{trivial_scatt}), 
the corresponding holomorphic vector bundle $E \ra \ov{T\bP^1}$ is a complex vector bundle of rank
$N$ with the structure group $U(N).$ \, Since $\ov{T\bP^1}$ is a 2-dimensional complex manifold,
the only non-vanishing Chern characters are the first and the second Chern characters, given by
\[ C_1(E) = \frac{i}{2 \pi} \mbox{Tr} F, \qquad   C_2(E) =  -\frac{1}{8 \pi^2} \mbox{Tr} (F \wedge F) \]
respectively, where $F$ is the curvature two-form of an arbitrary connection on $E.$   
(See for example \cite{Nakahara}.) 

The Chern number is an integer-valued topological invariant of $E,$ obtained by integrating the Chern characters over the base 
space $\ov{T\bP^1}.$  \, The condition that the bundle $E$ is trivial when restricted to the real sections of $\ov{T\bP^1}$
implies that the first Chern number vanishes.  To relate the second Chern number 
\[c_2(E) = \int_{\ov{T\bP^1}} C_2(E)\]
to the topological degree of the restricted extended solution $\psi$ given by (\ref{degree1}), we consider
the bundle   $E^* \ra {\mc F}$ over the restricted correspondence space, defined
by pulling back the bundle $E \ra \ov{T\bP^1}$ by the map $q: {\mc F} \ra \ov{T\bP^1}$ in (\ref{themap}),
i.e. $E^* :=q^* E.$  

An extended solution $\Psi(x,y,t, \ll)$ of the Lax pair (\ref{laxpair}) is a function on $\R^{2,1} \times \CP^1.$
Restricting $\Psi$ to the spacelike $t=0$ plane,  the matrix
\[ \psi_\ll(x,y,\ll)  := \Psi(x,y,0, \ll)\]
is a function on the finite region ${\mc R}=\R^2 \times \CP^1$ of the the restricted correspondence space ${\mc F}.$
The trivial scattering condition (\ref{trivial_scatt}) is that the restriction of $\psi_\ll$ to the real equator
$S^1 \subset \CP^1$ of the space of spectral parameter $\ll$ has the limit at spatial infinity
\[ \psi_\ll|_{S^1 \subset \CP^1} \equiv  \psi (x,y, \theta)
\longrightarrow \psi_0(\theta) \qquad \mbox{as} \qquad r = \sqrt{x^2+y^2}
\longrightarrow \infty, 
\] 
where $ \ll = -\cot(\frac{\theta}{2}).$ \, We can use a residual freedom in $\psi$ so that 
$ \psi (x,y, \theta) \longrightarrow {\bf I}$ as $r \ra \infty.$
\, The triviality of the vector bundle $E \ra \ov{T\bP^1}$ over $L_\infty$ guarantees that it is possible to choose
the extended solution $\psi_\ll$ such that
\be \label{psill} \psi_\ll (x,y, \ll) \longrightarrow {\bf I}  \qquad \mbox{as} \qquad r = \sqrt{x^2+y^2} \longrightarrow \infty. \ee 
This ensures that $\psi_\ll(x,y,\ll)$ extends to the region ${\mc R_\infty}$ and $e_{\tau_\R}$ of ${\mc F}.$

\vspace{1cm}

\noindent {\bf Proof of Theorem \ref{main_th}}

\vspace{0.3cm}

Let $E^*:=q^* E$ denote the pull back of the vector bundle $E \ra \ov{T\bP^1}$ by the map $q: {\mc F} \ra \ov{T\bP^1}$
defined in (\ref{themap}).  Then,
the second Chern number $c_2(E)$ of $E$ is given by 
\begin{eqnarray}
c_2(E) &=& \int_{\ov{T\bP^1}} C_2(E) = \int_{q({\mc F})} C_2(E)  \nonumber \\
&=&  \int_{\mc F} q^* C_2(E)  \label{qdiffeo} \\
&=&  \int_{\mc F} C_2( q^*E)  \label{qsmooth}  \\
&=&   c_2(E^*),  \nonumber
\end{eqnarray}
where we have used a property of integration of differential forms in (\ref{qdiffeo}), and that 
of the Chern characters in (\ref{qsmooth}).  Note that (\ref{qdiffeo}) is valid because the region
where $q$ is not a bijection is of codimension one, hence it does not contribute to the integral. \\

So now we reduce the problem to finding the second Chern number of $E^* \ra {\mc F}.$
\, Let ${\mc F_+}$ and ${\mc F_-}$ denote two open sets covering ${\mc F},$ given by
\[
{\mc F_+} :=  \{ (p,z) \in {\mc F}\; : \mbox{Im}(\ll) > -\vv \} \quad \mbox{and} \quad
{\mc F_-} :=  \{ (p,z) \in {\mc F}\; : \mbox{Im}(\ll) \; < \; \; \vv \},
\]
for some real constant $\vv > 0.$

The second Chern number  $c_2(E^*)$ is given by
\be  \label{FF} c_2(E^*) = -\frac{1}{8 \pi^2}  \int_{\mc F} \mbox{Tr} (F \wedge F), \ee
where $F$ is the curvature two-form of an arbitrary connection on $E^*.$ 
\, In the limit $\vv \ra 0,$ (\ref{FF})  becomes
\[ \label{FF1} c_2(E^*) = -\frac{1}{8 \pi^2}   \left[ \; \int_{\mc F_+} \mbox{Tr} (F \wedge F)  + \int_{\mc F_-}
\mbox{Tr} (F \wedge F) \; \right]. \]

Choose a connection one-form $A$ such that $A$ vanishes on ${\mc F_+}.$ \, Then, we only need to consider the 
integral over ${\mc F_-}.$
\, Now, since  $\mbox{Tr} (F \wedge F)$ is closed,  there exists a local three-form $Y$ such that
\[ \mbox{Tr} (F \wedge F) = d Y, \; \mbox{where} \; \;Y = \mbox{Tr}(F \wedge A -
\frac{1}{3} A \wedge A \wedge A).  \]
Then by Stokes' Theorem
\begin{eqnarray*} c_2(E^*) &=&  -\frac{1}{8 \pi^2} \int_{{\mc F_-}} d Y \;  = \; -\frac{1}{8 \pi^2} \int_{\p {\mc
      F_-}} Y \\
&=& -\frac{1}{8 \pi^2} \int_{\p {\mc F_-}}  \mbox{Tr}(F \wedge A - \frac{1}{3}
  {A}^3).
\end{eqnarray*}
The only boundary of ${\mc F_-}$ is the common boundary it has with
${\mc F_+}.$ \, Hence ${\p {\mc F_-} \subset {\mc F_+} \cap {\mc F_-}.}$ \\

Now, let $A_+$ and $A_-$ denote the connection $A$ in local trivialisations  over ${\mc F_+}$ and ${\mc F_-}$
respectively.   \, Then, over ${\mc F_+} \cap {\mc F_-}$
\[A_- \, = \, g^{-1}A_+ g + g^{-1} d g \, = \, g^{-1} d g, \]
since $A_+ \equiv 0,$ and where $g$ denotes the transition function of $E^*$ in the overlap.

Now the columns of $\psi_\ll$ can be used as meromorphic frame fields over ${\mc F_-}.$  
\, Note that $\psi_\ll$ is holomorphic and invertible  in the overlap ${\mc F_+} \cap {\mc F_-}$
for sufficiently small $\vv,$ because $\psi_\ll$ have poles only at non-real values of $\ll.$
\, In its own trivialisation, $\psi_\ll$ takes the form of the identity matrix. \, Now over ${\mc F_+},$
choose a local frame field such that $\psi_\ll$ takes the form $\psi_\ll(x,y,\ll)$ itself.
In these local trivialisations, the transition function $g$ is given by
\[ {\bf I} \, = \,g^{-1} \, \psi_\ll.  \]
Hence $ g \, = \, \psi_\ll. $ \,
Therefore, in the overlap ${\mc F_+} \cap {\mc F_-},$ the connection
 $A_-$ is given by
\[
A_- = \psi_\ll^{-1} d \psi_\ll.
\]

This implies that the curvature $F_-$ vanishes in ${\mc F_+} \cap {\mc F_-}$ and
\[ c_2(E^*) \, = \, \frac{1}{24 \pi^2}  \int_{\p {\mc F_-}}  {A_-}^3 \, = \, \frac{1}{24 \pi^2}  \int_{\p {\mc F_-}}
 \mbox{Tr}\left(({\psi_\ll^{-1} d \psi_\ll})^3\right).\]
Now, since over $\p{\mc F_-} \cap (e_{\tau_\R}
\cup {\mc R_\infty})$ the spacetime points $p$ belong to
$\RP^1_\infty,$  the boundary condition (\ref{psill}) implies that  
\be  A_- {\big|}_{ \p{\mc F_-} \cap (e_{\tau_\R} \cup {\mc R_\infty})} = 0, \ee
thus
\[ c_2(E^*) =    \frac{1}{24 \pi^2}  \int_{\p {\mc F_-} \cap {\mc R}}
 \mbox{Tr}\left(({\psi_\ll^{-1} d \psi_\ll})^3\right). 
\]
Then, since 
\, $ \p {\mc F_-} =  \{ (p,z) \in {\mc F}\; : \mbox{Im}(\ll) =0 \},  $
\, i.e. $\ll \in S^1 \subset \CP^1,$  we have
\[\p {\mc F_-} \cap {\mc R} = \R^2 \times S^1 \qquad \mbox{and} \qquad \psi_\ll\big|_{\p {\mc F_-} \cap {\mc R}} = \psi(x,y,\theta).   \]
Therefore 
\[ c_2(E^*) =  \frac{1}{24 \pi^2}  \int_{\R^2 \times S^1}  \mbox{Tr}
\left(({\psi^{-1} \d \psi})^3\right)=  \frac{1}{24 \pi^2}  \int_{S^3}  \mbox{Tr}
\left(({\psi^{-1} \d \psi})^3\right) = [\psi] \]
where we have used the fact the $\psi$ satisfies the trivial scattering condition, thus its domain extends to $S^3.$
\koniec

\vspace{0.5cm}

\noindent {\bf Remark.}  Theorem \ref{main_th} holds for any Ward chiral field which 
satisfies the finite energy condition (\ref{assympt}) and the trivial scattering condition (\ref{trivial_scatt}).
A subclass of such solutions are the so-called time-dependent $n$-unitons, where the extended solutions $\Psi(x,y,t,\ll)$ have 
a pole of order $n$ at $\ll = \mu,$ where  $\mu \in \C \backslash \R.$  The total energy of a time-dependent uniton is known  \cite{DP06} to be
directly proportional to $[\psi].$  Hence, we deduce that the total energy of a time-dependent uniton is 
proportional to $c_2(E),$ consistent with the result for statistic Ward chiral fields in \cite{An0}.

\section*{Acknowledgements}
Section \ref{sec:compminitwist} and \ref{sec:corresspc} of this paper are parts of the author's
PhD Thesis ``Anti-Self-Duality and Solitons" under the supervision of Maciej Dunajski.  
The author wishes to thank Maciej Dunajski and Lionel
Mason for valuable comments  on aspects of the compactified twistor fibration.
The author is also grateful to the Faculty of Science, Prince of Songkla University for research support.

\vspace{1cm}

\renewcommand\appendix{\par
  \setcounter{section}{0}
  \setcounter{subsection}{0}
  \setcounter{figure}{0}
  \setcounter{table}{0}
 \setcounter{equation}{0}
\setcounter{theo}{0}
  \renewcommand\thesection{Appendix \Alph{section}}
 \renewcommand\theequation{\Alph{section}\arabic{equation}}
 \renewcommand\thetheo{\Alph{section}\arabic{theo}}
}

\appendix

\section{Null planes in $\C^3$ and Ward correspondence} \label{appa}
\setcounter{equation}{0}

 A null plane in a complexified spacetime $M_\C = \C^3,$ with the flat metric
\be \label{c3flatmetric} ds^2 = -dt^2 + dx^2 +dy^2, \ee 
is defined  
 as  a 2-plane whose
normal vector is null with
respect to the metric (\ref{c3flatmetric}).  Hence, the equation for a null plane is given by
\be \label{nullplaneequation} \eta_{\mu \nu} k^\mu x^\nu = -\frac{1}{2}\h \om,   \ee 
where $x^\mu = (x^0 =t, \; x^1 = x, \; x^2 = y),$ 
$\eta_{\mu \nu}= \mbox{diag}(-1, 1, 1),$ $k^\mu$ is the normal null vector field
and $\h \om$ is a constant.  The factor of $-\frac{1}{2}$ is
 introduced for convenience, the reason for 
which will become apparent shortly.  To parametrise null vector fields, it
is  useful to use the spinor
formalism based on the identification 
\[ TM_\C = {\mc S} \odot {\mc S}, \]
where $TM_\C$ is the holomorphic tangent bundle of $M_\C$ and ${\mc S}$ is a rank two vector bundle over $M_\C.$   A tangent vector field
$k$ can  be written as a symmetric two-spinor
\[ k^{AB} =  \left( \begin{array}{cc}
    k^0 + k^2 & k^1 \\
    k^1 & k^0 - k^2 
  \end{array}
  \right),  \]
such that $\eta_{\mu \nu} k^\mu k^\nu = - \det( k^{AB}) = -\frac{1}{2}
\vv_{AC} \vv_{BD} k^{AB} k^{CD}.$  It follows  that a 
 null vector field corresponds to a symmetric two-spinor of rank 1.
 That is, every null vector field is given by
$k^{AB} = \pi^A \pi^B$ for $\pi^A \ne (0, 0),$ where $\pi^A, \; A =0,1,$ denotes the fibre
coordinates of ${\mc S}.$   Now, writing the spacetime coordinates also 
as a symmetric two-spinor
\[ x^{AB} = \left( \begin{array}{cc}
    t + y & x \\
    x & t - y
  \end{array}
  \right),  \]
 the null plane equation (\ref{nullplaneequation}) becomes
\be \label{nullplanehomo2} \h \om = x^{AB} \pi_A \pi_B.  \ee
Moreover, let us assume that $\pi_1 \ne 0.$  \, Defining $\om = \h \om/ \pi_1^2$ and $\dsl \ll = \frac{\pi_0}{\pi_1},$
 equation (\ref{nullplanehomo2}) now reads 
\be \label{nullplaneeq2} \om = (t+y) \ll^2 + 2x \ll +(t-y). \ee
  The null planes with
 $\pi_1 = 0$ can also be captured by (\ref{nullplaneeq2}) by allowing
 $\ll$ to go to infinity.  This implies that every null plane in $\C^3$ is
 labelled by $(\om, \ll),$ where $\om \in \C$ and $\ll \in \CP^1.$
 The minitwistor space, which is the space of null planes in $M_\C=\C^3,$ is
 therefore a line bundle over $\CP^1.$  It is in fact the tangent
 bundle $T\bP^1$ of $\CP^1.$  To see this, note that under the change
 of coordinate $\ll \ra \t \ll = \ll^{-1},$ the fibre
 coordinate changes by $\om \ra \t \om = \om \ll^{-2}.$  \\

It follows
 from (\ref{nullplaneeq2}) that a point $p \in M_\C$ corresponds to a holomorphic section 
 $ \h p$ of $ T\bP^1.$  We can define
the correspondence space ${\tt F}$ to be the space of pairs $(p, Z),$ of a point
$p \in M_\C$ and a null plane $Z$ passing through $p.$   There is a $\CP^1$ worth of
null planes passing through each point $p \in M_\C,$ and thus
${\tt F} =  \C^3 \times \CP^1.$
  Note that the
two vector fields in the Lax pair (\ref{laxpair})
\[ l_0 = \p_x - \ll (\p_t - \p_y), \qquad l_1 = \p_t + \p_y - \ll \p_x
\]
span a null plane, as they annihilate $\om = (t+y) \ll^2 + 2x \ll
+(t-y).$  The minitwistor space $T\bP^1$ can therefore be regarded as the quotient
space of $\C^3 \times \CP^1$ by the distribution $\{l_0, l_1\}.$
 
Since the Ward chiral model (\ref{Wardeq}) is the compatibility condition of
the Lax pair (\ref{laxpair}), there exist $N$ linearly independent column vector
solutions of (\ref{laxpair}) if $J$ in (\ref{gaugechoice}) is a solution of
(\ref{Wardeq}).  These column vector solutions are the covariantly
constant sections with respect to $(A, \Phi),$  of the trivial
$\C^N$ bundle $V \ra M_\C$ restricted to null planes.
One can construct a holomorphic  rank $N$ vector bundle over the
minitwistor space $T\bP^1$ by taking the fibre over each point $z \in
T\bP^1$ to be the space of covariantly constant sections of $V|_Z,$ where $Z$
is the null plane corresponding to the point $z \in T\bP^1.$ \\


\section{The blow-up of the cone} \label{appb}
\setcounter{equation}{0}

 A reference for the blow-up can
be found, for example, in \cite{GH78, Hartshorne}.
First, let us consider the blow-up of an open set $U = \C^3
\subset \CP^{3*}$ 
with $Z_0 \ne 0.$   Let 
\be \label{z1z2z3coord} z_1= \frac{Z_1}{Z_0}, \quad z_2= \frac{Z_2}{Z_0}, \quad z_3=
\frac{Z_3}{Z_0}\ee
be coordinates on $U,$ so that the vertex ${\bf z_0}$
coincides with the origin.
By definition, the blow-up $\wt U$
of $U$ at the origin is  given by
\be \label{blowup0} 
\{ (z,l) \in U \times \CP^2 : z_il_j = z_jl_i, \; i \ne j \}, 
\ee
where $\{l_i\}$ are  homogeneous coordinates of the $\CP^2,$ $i=1,2,3.$
 In other words, $\wt U$ is a $3$-dimensional subspace of $U
\times \CP^2$  defined by the relation in
(\ref{blowup0}).  Geometrically, $z$ lies on a line labelled by $l \in \CP^2$
passing through the origin
in $\C^3.$   One can consider $\wt U$ in three coordinate neighbourhoods:
 $\wt U^k$, where $l_k \ne
0.$   
A point in $\wt U^k$ is labelled by  $(z_k,
\frac{l_j}{l_k})$ where  $j \ne k.$

There exists a surjective map from
$\wt U$ to $U$ which is given locally in a coordinate patch $\wt U^k$ by
\be \label{projectionmap} \pi: \left( z_k, \frac{l_j}{l_k} \right)
\mapsto \left( z_k, \; z_j = z_k \frac{l_j}{l_k} \right).\ee
One sees that for a point $(z_k, z_j)$ with $z_k \ne 0$ there is a
unique preimage $\left( z_k,\; \frac{l_j}{l_k} = \frac{z_j}{z_k}
\right)$ in $\wt U.$   However, if
$z_k =0$  all points with coordinates 
$\left(0,\frac{l_j}{l_i} \right)$ are mapped to the origin.  
Hence, the preimage ${\tt E}$ of the origin is isomorphic to $\CP^2.$
The set ${\tt E}$ is called  
the exceptional divisor.   It is important to note that the map 
$\pi: \wt U - {\tt E} \lra U-\{ {\bf z_0} \}$ is one-to-one. \\ 

Now, let us look at the blow-up $\wt {\mc C}$ at the vertex of the cone ${\mc C}
\subset \CP^{3*}.$  We are only interested in the region around
the vertex, as the projection from $\wt {\mc C}$ to ${\mc C}$ is $1:1$
elsewhere.  The blow-up $\wt {\mc C}_U = \wt {\mc C} \cap \wt U$ is
obtained from  $\wt U$ by imposing the cone equation 
(\ref{homocone}) on $\wt U.$  \, In a 
 coordinate patch, say $\wt U^1$ where $l_1 \ne 0,$ (\ref{homocone}) becomes
\[
z_1^2 \left( 1+ \left( \frac{l_2}{l_1} \right)^2 - \left(
\frac{l_3}{l_1} \right)^2  \right) = 0.
\]
 The continuity implies that $\wt {\mc C}_U \cap
{\tt E}$ are given locally in $\wt U^1$ by the points with 
\[ z_1
=0 \quad \mbox{and} \quad  1+ \left( \frac{l_2}{l_1} \right)^2 - \left(
\frac{l_3}{l_1} \right)^2 =0.\]
  Since $l_1 \ne 0$ in $\wt U^1,$ the second
condition can be written as 
\be
 \label{blowvertex} l_1^2+l_2^2-l_3^2 = 0. 
\ee
 One obtains similar
descriptions in patches $\wt U^2$ and $\wt U^3.$  Then it follows from 
(\ref{blowvertex}) that $\wt {\mc C}_U \cap {\tt E} =\wt {\mc C} \cap {\tt E} $ is $\{ {\bf z_0} \}
 \times \CP^1,$ where the $\CP^1$ is embedded in the $\CP^2 \ni [l_i]$  by 
\be \label{alpha} 
[l_1, l_2, l_3] = [-2 \pi_0 \pi_1, \; {\pi_1}^2 - {\pi_0}^2, \; {\pi_0}^2+{\pi_1}^2 ]
\ee
with $\pi_A \in \C^2-\{0\},$ where we have used the same parametrisation as for  null
 vectors.  The $\CP^1$ in (\ref{alpha}) can be parametrised by a single
variable as 
\be \label{paraml}  [ -2\ll, 1-\ll^2, 1+\ll^2 ] \qquad \mbox{and} \qquad [2\t \ll, {\t \ll}^2-1,
  1+{\t \ll}^2],    \ee
in the patches with $\pi_1 \ne 0$ and $\pi_0 \ne 0$ respectively, where
$\t \ll = \frac{1}{\ll}$ in the overlap.

Note that we deliberately denote the inhomogeneous coordinate of the
$\CP^1$ by  $\ll,$ to be the same as
the base coordinate of $T\bP^1.$  We shall now show 
that the $\wt {\mc C} \cap {\tt E}$ indeed corresponds to  $L_\infty$ - the additional
$\CP^1$ at $\om = \infty$ of $\ov {T\bP^1}.$ \\

\noindent {\bf Proof of Proposition \ref{blowz0andLinfty}.}
A bijection from $T\bP^1$ to ${\mc C} - z_0$ is already given by  
(\ref{map1}).  Here, we shall extend the map (\ref{map1}) to a
bijection from  $\ov {T\bP^1}$ to $\wt {\mc C}.$  
Although the fibre of $\ov {T\bP^1}$ is a $\CP^1,$ we shall avoid
using two fibre-coordinate patches, but rather we will define
a map by taking the limit $\om \ra \infty.$   Since the projection
$\pi: \wt {\mc C} - (\wt {\mc C} \cap {\tt E}) \lra {\mc C}- \{{\bf z_0}\}$
is one-to-one, we only need to consider the map locally in a neighbourhood
of ${\bf z_0}.$   
Assuming $\om \ne 0,$ then (\ref{map1}) can be written as 
\be \label{map2}
(\om , \ll) \mapsto [1, \frac{-2\ll}{\om},\frac{1-\ll^2}{\om},
  \frac{1+\ll^2}{\om} ], \quad  \mbox{and} \quad
(\tilde \om , \tilde \ll) \mapsto [1, \frac{-2\t \ll}{\t \om},
    \frac{{\t \ll}^2-1}{\t \om},
  \frac{1+{\t \ll}^2}{\t \om} ].
\ee

To extend the domain of (\ref{map2}) to $\ov {T\bP^1}$ minus the $\om
= 0$ section, we shall take
the limit $\om \ra \infty.$  For concreteness, let us consider the
first local map of (\ref{map2}).  In the inhomogeneus coordinates
$z_1, z_2, z_3$  (\ref{z1z2z3coord}) of  
${\mc C}_U - \{{\bf z_0}\},$ the first map of (\ref{map2}) is given by 
\be \label{map3}
(\om \ne 0, \ll) \longmapsto (z_1,
z_2,z_3)=\left( \frac{-2\ll}{\om},\frac{1-\ll^2}{\om},
\frac{1+\ll^2}{\om} \right).
\ee

We can now define another map from the image of (\ref{map3}), which
is ${{\mc C}_U -\{{\bf z_0}\},}$ to the blow-up $\wt {\mc C}_U$ in terms of three local maps from
the three regions ${U^1=\{ \ll \ne 0\},}$
$U^2=\{ \ll \ne \pm 1\}$ and $U^3=\{ \ll \ne \pm i\}$ to the blow-up 
neighbourhoods $\wt U^1 = \{ l_1 \ne 0 \},$ $ \wt U^2 = \{ l_2 \ne 0
\}$ and $\wt U^3 = \{ l_3 \ne 0 \}$ in $\wt U$ respectively. \\

In $U^1$ for example, the local map is defined by
\[
(z_1,z_2,z_3) \lmt \left( z_1, \; \frac{l_2}{l_1} = \frac{z_2}{z_1}, \;
  \frac{l_3}{l_1} = \frac{z_3}{z_1} \right). \]
Composing it with the map (\ref{map3}), we have
\be \label{map4} (\om \ne 0, \ll) \lmt \left( z_1, \frac{l_2}{l_1},
  \frac{l_3}{l_1} \right)= 
 \left( \frac{-2\ll}{\om},\frac{1-\ll^2}{-2\ll}, \frac{1+\ll^2}{-2\ll}
  \right). 
\ee
One sees that this is consistent with the parametrisation of $[l_i]$ in (\ref{paraml}).
Since at this point $\om$ is still finite and $\ll \ne 0$ in $U^1,$
 then  $z_1 \ne 0.$  
Therefore (\ref{map4}) is a
one-to-one map from $U^1$ to $\wt U^1,$ whose image is 
$ (\wt {\mc C} \cap \wt U^1) - (\wt U^1 \cap {\tt E}).$
The local maps  $U^2 \ra \wt U^2$ and $U^3 \ra \wt U^3$ are defined
similarly from the inverse of the projection (\ref{projectionmap}).

Now, consider the limit  $\om \ra \infty$ of the map (\ref{map4})
\be \label{map4limit} (\om \ne 0, \ll) \longmapsto \left( z_1, \frac{l_2}{l_1},
  \frac{l_3}{l_1} \right)=
  \left(0,\frac{1-\ll^2}{-2\ll},\frac{1+\ll^2}{-2\ll} \right) \; \; \mbox{as}
\; \om \ra \infty.
\ee
We define a bijection from $\ov {T\bP^1} \cap U^1 $ to $ \wt {\mc C} \cap \wt U^1$ to be the
 extension of the map (\ref{map4}) by the limit (\ref{map4limit}).
Comparing this with the local expression of  $\wt {\mc C} 
\cap {\tt E} \cap \wt U^1$ obtained from  (\ref{paraml}), and similarly for the other two
neighbourhoods, one deduces that $L_\infty$ is mapped onto the restricted exceptional divisor
$\wt {\mc C} \cap {\tt E}.$ \koniec  

\vspace{0.5cm}

\noindent {\bf Remark.} 
 Since every map is given in holomorphic
coordinates, one deduces that $\ov {T\bP^1}$ is biholomorphic to $\wt
{\mc C}.$ 

\vspace{0.5cm}

\section{The correspondence space} \label{appc}
\setcounter{equation}{0}

We define the correspondence space  $\hat {\mc F}$ of a double fibration
to the compactified spacetime $\ov {M_\C} = \CP^3$ and the
compactified twistor space $\ov{T\bP^1} \cong \wt {\mc C}$ to be 
 the blow-up of the algebraic variety $\hat f $
  (\ref{algebraic}) along $\CP^2_\infty \times \{ {\bf  z_0} \}.$ \\

\noindent The correspondence space $\hat {\mc F}$ has the following properties. 

\vspace{0.5cm}

\noindent {\bf 1.}  {\bf The blow-up of $\CP^2_\infty \times \{ {\bf
  z_0} \}$ is $\CP^2_\infty \times L_\infty.$}

   This  can  be derived from the direct construction of
 the blow-up as follows.
\, Let us first consider the blow-up of $\CP^3 \times \CP^{3*}$ along $\CP^2_\infty \times \{ {\bf
  z_0} \}$ locally in each coordinate patch.  Recall
  that $\CP^2_\infty = \{ [P^\a] : P^0=0 \}$ and ${\bf
  z_0} = [1,0,0,0].$  Since we know that away from the singularity, the
projection $\rho: \hat {\mc F} \lra \hat f$ is a $1:1$ and onto, 
we need to consider only three coordinate patches of $\CP^3 \times
  \CP^{3*}$ that include the
singularity, namely $U_i = \{ {Z_0 \ne 0,} \; P^i \ne 0\}, \; {i = 1,2,3.}$ 
Then, the blow-up of $\hat f \subset \CP^3 \times \CP^{3*}$ is obtained by imposing the incidence
  relations in  (\ref{algebraic}).   Note the lower index of $U_i,$
  to be distinguished from $U^i$ in \ref{appb}. 

 First,
  consider the patch $U_1 = \{ Z_0 \ne 0, P^1 \ne 0\} = \C^3
  \times \C^3 =\C^6$ with  
 coordinates
\[ (y_i) = ( y_0 = p^0,\; y_1 =z_1,\; y_2 = z_2,\; y_3 = z_3,\;  y_4 =
p^2,\;  y_5 = p^3 ),\]
 where $z_j =
\frac{Z_j}{Z_0}$ and $p^j=\frac{P^j}{P^1}.$ 
   The 
intersection $ (\CP^2_\infty \times \{ {\bf  z_0} \}) \cap U_1 $ is
 then given by $\C^2_{\infty (1)} := \{(0,0,0,0,p^2, p^3)\} =
 \C^2.$  The blow-up of 
$U_1$ along $\C^2_{\infty (1)}$ is by definition (see for example \cite{GH78}) given by
\be \label{blowU_1}
\wt U_1 := \{ (y,l) \in \C^6 \times \CP^3 \; : y_il_j = y_jl_i, \;
i\ne j \in  \{0,1,2,3\} \}, \ee
where $\{l_i\}$ are homogeneous coordinates of the $\CP^3.$  \, The
projection ${\rho: (y,l) \mapsto y}$ is bijective onto the region away
from  $\C^2_{\infty (1)}.$    If $y \in \C^2_{\infty (1)},$ then $l$ is
arbitrary, and
hence the preimage of $\C^2_{\infty (1)}$ is $\hat {\tt E}_1 :=
\rho^{-1}(\C^2_{\infty (1)}) =
\C^2_{\infty (1)} \times \CP^3.$  

The blow-up $\hat {\mc F} \cap \wt U_1 $ is obtained from $\wt U_1$ by
imposing the incidence relations in (\ref{algebraic})
\be \label{alg1} z_1^2+z_2^2-z_3^2=0, \;\; \mbox{and} \;\; p^0+z_1 + p^2z_2-p^3z_3 =0.\ee
Lifting the relations  (\ref{alg1}) to $\wt U_1,$ they can be
written locally in the four coordinate neighbourhoods of $\wt U_1.$
First, in the patch $l_0 \ne 0,$  with the coordinates $(p^0, \frac{l_1}{l_0},
\frac{l_2}{l_0},\frac{l_3}{l_0},p^2, p^3),$ equation (\ref{alg1}) becomes
\be \label{alg1.1}
(p^0)^2 \left( \left(\frac{l_1}{l_0} \right)^2
+\left(\frac{l_2}{l_0} \right)^2 - \left(\frac{l_3}{l_0} \right)^2
\right) = 0 \quad \mbox{and} \quad
p^0 \left( 1+ \frac{l_1}{l_0} + \frac{l_2}{l_0} p^2 - \frac{l_3}{l_0}
p^3 \right) = 0. 
\ee

Recall that the exceptional divisor of $\wt U_1$ is given by
\[\hat {\tt E}_1 = \{(y,l) \in \wt U_1 \; : y_0=y_1=y_2=y_3=0 \}, \]
which is the preimage of $\C^2_{\infty (1)}.$ 
 The continuity of
(\ref{alg1.1}) implies that 
$\hat {\mc F} \cap \hat {\tt E}_1$ is given locally in the patch $l_0 \ne 0$
by the points with $p^0=0$  and
\[
\left(\frac{l_1}{l_0} \right)^2
+\left(\frac{l_2}{l_0} \right)^2 - \left(\frac{l_3}{l_0} \right)^2
 = 0, \qquad 
1+ \frac{l_1}{l_0} + \frac{l_2}{l_0} p^2 - \frac{l_3}{l_0}
p^3  = 0.
\]
Since $l_0 \ne 0,$ the last two equations can be written as 
\begin{eqnarray} \label{alg1.2}
l_1^2+l_2^2-l_3^3 &=& 0 \\
\label{alg1.3}
l_0+l_1+l_2p^2-l_3p^3 &=& 0.
\end{eqnarray}
Similarly, in the patch $l_1 \ne 0,$ with coordinates
$(\frac{l_0}{l_1}, z_1,\frac{l_2}{l_1},\frac{l_3}{l_1},p^2, p^3),$
equation (\ref{alg1}) becomes 
\[
(z_1)^2 \left( 1+\left(\frac{l_2}{l_1} \right)^2 -
 \left(\frac{l_3}{l_1} \right)^2 \right)
 = 0, \qquad
z_1 \left( \frac{l_0}{l_1} + 1+ \frac{l_2}{l_1} p^2 - \frac{l_3}{l_1}
p^3 \right) = 0,  
\]
and $\hat {\mc F} \cap \hat {\tt E}_1$ is locally given in this
neighbourhood by the points with $z_1 =0$ and 
\[
1+\left(\frac{l_2}{l_1} \right)^2 - \left(\frac{l_3}{l_1} \right)^2
 = 0 \quad \mbox{and} \quad
\frac{l_0}{l_1} + 1+ \frac{l_2}{l_1} p^2 - \frac{l_3}{l_1}
p^3  = 0.
\]
Now, since $l_1 \ne 0,$ the above equations can also be written in
the homogeneous coordinates $\{l_i\}$ as (\ref{alg1.2}, \ref{alg1.3}).
The equations in the other two patches $l_2\ne 0$ and $l_3 \ne 0$
are similar to those in the patch $l_1 \ne 0.$  

\vspace{0.3cm}

This shows that $\hat {\mc F} \cap \hat {\tt E}_1$ is the subset of 
\[\hat {\tt E}_1 = \{ (0,0,0,0,p^2,p^3,l_0,l_1,l_2,l_3) \} = (\C^2_{\infty (1)}
\times \CP^3) \subset (\C^6 \times \CP^3)\]  
 given by (\ref{alg1.2}, \ref{alg1.3}).  Now, given $(p^2,p^3),$  
  $l_0$ is uniquely determined from $(l_1,l_2,l_3)$ by
 (\ref{alg1.3}).  This, together with (\ref{alg1.2}), defines a $\CP^1
 \subset \CP^3$  given by 
\[ [l_0, \; -2 \a_0 \a_1, \; \a_1^2 - \a_0^2, \; \a_0^2+\a_1^2],  \]
where $l_0 = 2 \a_0 \a_1  + (\a_0^2 - \a_1^2)p^2 +
(\a_0^2+\a_1^2)p^3$ and $\a_A \in \C^2 -\{0\}.$  Hence, we conclude that $\hat {\mc F} \cap \hat{\tt E}_1 =
  \C^2_{\infty (1)} \times \CP^1.$   Note that the local
equations for $\hat {\mc F} \cap \hat {\tt E}_1$ are smooth and in fact
holomorphic in each of the four patches of $\wt U_1.$ \\

In the other two patches \, $U_2 = \{ Z_0 \ne 0, P^2 \ne 0\}$ and ${U_3 =
\{ Z_0 \ne 0, P^3 \ne 0\},}$  the blow-up follows similarly.   Let
$\hat {\tt E} = \hat {\tt E}_1 \cup \hat {\tt E}_2 \cup \hat {\tt E}_3$
 denotes the union of
the exceptional divisors of $\wt U_1, \wt U_2$ and $\wt U_3.$  The
blow-up is defined such that the coordinate patches glue
naturally, therefore we conclude that $\hat {\mc F} \cap \hat {\tt E} =  \CP^2_\infty
\times \CP^1.$   Note that since $\{U_i\},$ $i = 1, 2, 3,$ are the only patches that include the
singularity  $\CP^2_\infty \times \{{\bf z_0}\},$ then $\hat {\tt E}$ is also the
exceptional divisor of the blow-up of $\CP^3 \times \CP^{3*}$ along
$\CP^2_\infty \times \{{\bf z_0}\}.$  

The $\CP^1$ in $\hat {\mc F} \cap \hat {\tt E}$ is precisely
$L_\infty$ of $\wt {\mc C}.$  To see this, consider
 the incidence relation (\ref{blowU_1}) in $\wt U_1.$  Equation $y_il_j=y_jl_i$ 
 implies  $z_il_j=z_jl_i, \; i=1,2,3,$ and the same
holds for $\wt U_2$ and $\wt U_3.$  This is the same expression for the blow
up of ${\mc C}$ along ${\bf z_0}.$   \\

\noindent {\bf 2.}  {\bf $\h {\mc F}$ is a $\CP^2$ bundle over $\wt {\mc
    C}.$}

This feature gives another direct way to show that $\h {\mc F} \cap
\h {\tt E} = \CP^2_\infty \times L_\infty.$
Let us start with the fact that the algebraic variety 
$\hat f \subset \CP^3
\times {\mc C}$ given by  (\ref{algebraic}) is a $\CP^2$ 
bundle over ${\mc C}.$ To see this, consider the followings.
 Let us  denote the neighbourhood $\{ Z_0 \ne 0\} \subset \CP^3
\times \CP^{3*}$ by $U,$ the same as the neighbourhood $\{ Z_0 \ne 0\} \subset \CP^{3*}.$ 
\, Locally in $U,$ with coordinates $z_i = \frac{Z_i}{Z_0},$ the
incidence relations in (\ref{algebraic}) become
\begin{eqnarray}  \label{Vcone} z_1^2 + z_2^2-z_3^2 &=& 0 \\ 
         \label{Vplane}    P^0  + P^1z_1 + P^2z_2 - P^3z_3 &=& 0.   
\end{eqnarray}
 Equation (\ref{Vplane})
is homogeneous, i.e. given a point $(z_1,z_2,z_3)$ on
${\mc C}_U$ satisfying (\ref{Vcone}), a solution
$[P^\a] \in \CP^3$ is given by 
\[ [-P^1z_1 - P^2z_2 +P^3z_3, P^1, P^2,P^3 ], \]
which is determined by $(P^1,P^2,P^3)$ up to a constant 
multiplication.  This implies that $\hat f_U := \hat f \cap U =  \CP^2_\infty \times {\mc
  C}_U,$ where $\CP^2_\infty$
is the $\CP^2$ corresponding to ${\bf z_0}.$  
In the other neighbourhoods of ${\mc C},$ for example $W=\{ z \in {\mc
 C} : Z_1 \ne
 0\},$ the subset $\hat f_W := \hat f \cap W$ is also biholomorphic to
 $\CP^2 \times {\mc C}_W,$ where in this case the $\CP^2$ is given by $[P^0,
 \frac{(-P^0Z_0 - P^2Z_2 +P^3Z_3)}{Z_1}, P^2,P^3 ].$  This shows that
 $\h f$ is a $\CP^2$ bundle over ${\mc C}.$

\vspace{0.5cm}

Now, $\h{\mc F}$ is the blow-up of $\h f$ along $\CP^2_\infty \times {\bf
  z_0}.$ 
Locally the blow-up $\h {\mc F}_U$ is biholomorphic to $ \CP^2 \times \wt {\mc C}_U,$ where $\wt
{\mc C}_U$ is the blow-up of ${\mc C}_U$ along ${\bf z_0}.$   
To see this, consider the blow-up locally in the three regions of
$\hat f_U,$ namely $U_1 = \{ Z_0 \ne 0, P^1 \ne 0\},$ $U_2 = {\{ Z_0
\ne 0,} {P^2 \ne 0\}}$ and $U_3 = \{ Z_0 \ne 0, P^3 \ne 0\}$ as
discussed previously.  Note that $\hat f_U$ is
completely covered by these three open sets.  The points in $U$ which 
are omitted by $U_1, U_2, U_3$ are the ones with $(P^1,P^2,P^3) =
(0,0,0)$ and  are not solutions of (\ref{Vplane}). 

We have already done the blow-up $\h {\mc F}_{U_1}$ of $\hat f_{U_1}$ explicitly, where we
describe it in coordinate patches, $\{l_0 \ne 0\}, \; \{l_1 \ne 0\},\;
\{l_2 \ne 0\},$ and $\{l_3 \ne 0\}.$  However, we note here that $\hat {\mc
  F}_{U_1}$ can in fact be described completely in the patches  $\{l_i \ne
0\}, \; i=1,2,3,$ because the point $(l_0 \ne 0, 0,0,0)$ is not a solution
of (\ref{alg1.3}).  In the patch $l_1 \ne 0,$ with coordinates $(\frac{l_0}{l_1},
z_1,\frac{l_2}{l_1},\frac{l_3}{l_1},p^2, p^3)$ we can label  a point
in $\hat {\mc F}_{U_1, \; l_1 \ne 0}$ by
\[ (-1-\frac{l_2}{l_1}p^2 + \frac{l_3}{l_1}p^3, \;
z_1,\; \frac{l_2}{l_1}, \; \frac{l_3}{l_1},\; p^2, \; p^3), \]
as a consequence of (\ref{alg1.3}).  The set $(z_1, \frac{l_2}{l_1},\frac{l_3}{l_1} )$ can be identified with a
point in $\wt {\mc C}_U.$
Hence, given a point $z \in \wt {\mc C}_U$ we  only have freedom in
$(p^2,p^3).$  Let $\C^2_{(1)}$ denote the $\C^2$ defined by $(p^2, p^3).$ 
Then, we have that $\hat {\mc F}_{U_1 \; l_1 \ne 0} = \C^2_{(1)} \times \wt {\mc
  C}_U.$  One can deduce the same result for the patches $l_2 \ne 0,$ $l_3
\ne 0,$ and therefore $ \hat {\mc F}_{U_1} = \CP^2 \times \wt {\mc C}_{U_1}.$
This, together with similar results from the neighbourhood $U_2$ and
$U_3,$ imply that  
\[\hat {\mc F_{U}} = \CP^2 \times \wt {\mc C}_{U},\]
where it follows that $\hat {\mc F} \cap \hat {\tt E} =\CP^2_\infty
\times L_\infty.$   Moreover, since $\h {\mc F} - \h {\mc F_U}$ is
biholomorphic to $\h f - \h f_U,$ we conclude that 
$\h {\mc F}$ is a $\CP^2$ bundle over $\wt {\mc C}.$ \\


\section{The restricted double fibration} \label{appd}
\setcounter{equation}{0}

The restricted correspondence space  ${\mc F}$ defined in Section \ref{sec:restcorres} admits a surjective map to $\ov {T\bP^1}.$  This is due
to the following proposition.

\begin{prop} \label{intersectnullplaneprop}
Let $(P^0, P^1, P^2, P^3) \in \C^4-\{0\}$ be homogeneous
coordinates of a compactified complexified spacetime  $\ov {M_\C} =
\CP^3,$ and let $\tau_\R$ denote an $\RP^2 \subset
\ov {M_\C}$ defined by $(P^0, P^1, P^2, P^3) \in \R^4-\{0\}$ and 
$P^3=0.$  Recall that a null plane in $\ov {M_\C}$ is defined to be a $\CP^2$ given by
\[ Z_0P^0 +Z_1P^1+Z_2P^2 - Z_3P^3 = 0, \]
with $(Z_0, Z_1, Z_2, Z_3 ) \in \C^4-\{0\}$ satisfying
$ ({Z_1})^2 + ({Z_2})^2 - ({Z_3})^2 = 0.$

Then, every null plane in $\ov {M_\C}$ intersects $\tau_\R.$
\end{prop} 

\vspace{0.5cm}

\noindent {\bf Proof of Proposition  \ref{intersectnullplaneprop}.}
Let us first consider the class of real null planes. 
Let $\RP^{3*}$ be the subset of $\CP^{3*}$ that consists of points
$[Z_\a]$ whose representatives can be chosen to be in $\R^4-\{0\},$
and let ${\mc C}_{\R}$ be the intersection ${\mc C} \cap \RP^{3*}.$
We call the planes corresponding to $z \in {\mc C}_\R$ real null planes.
To see the intersection of real null planes with $\tau_\R,$ we first 
look at the real null planes  with $Z_0 \ne 0.$ 
 Such a null plane is given by 
\be \label{P0}  P^0+P^1z_1 + P^2z_2 =0,\ee
with $(z_1,z_2,z_3)$ all real.
Since everything is real, given  $(z_1,z_2,z_3),$ 
$P^0$ is determined in terms of $P^1, P^2$ by (\ref{P0}). Thus, the
intersection of a null plane with $Z_0 \ne 0$ with $\tau_\R$ is an $\RP^1.$  For a real null
plane with $Z_0=0,$ either $Z_1$ or $Z_2$ must be non-zero. 
Similar calculation for these planes shows that their intersections
with $\tau_\R$ are also $\RP^1.$
Therefore, one concludes that each real null plane intersections $\tau_\R$ in an $\RP^1.$   
Note that $\CP^2_\infty$ is a real null plane, whose 
intersection with $\tau_\R$ is $\RP^1_\infty.$  \\

Let us call the real null
planes with $(Z_1, Z_2, Z_3) \ne (0,0,0)$ {\it finite real null planes}.
For finite real null planes we have
the parametrisation (\ref{homoparamcone}), where the coordinates $[\hat
  \om, \pi_A]$ can be chosen such that $\hat \om \in \R,
\; \pi_A \in \R^2 - \{0\}.$   We will now show that these planes
intersect $\tau_{\R}$ in oriented
lines which are the extension of straight lines in $t=0$ $\R^2$-plane.  
First, note that  a finite real null plane  corresponds to a null plane
in $\R^{2,1} $ given by
\[\hat \om = 2x \pi_0 \pi_1 + y (\pi_0^2 - \pi_1^2) + t (\pi_0^2 +
\pi_1^2),\]
where we use $(x, y, t)$ in (\ref{xytcoord}) as  coordinates on $\R^{2,1}.$
Using the diffeomorphism between $\RP^1$ and
$S^1$
\[ \pi_0 = \cos \left( -\frac{\theta}{2} \right), \quad \pi_1 = \sin
\left( -\frac{\theta}{2} \right), \]
the null plane equation becomes
\be \label{R21nullplane}
 \hat \om = t - x \sin \theta + y \cos \theta. \ee
Now, the intersection with $\tau_\R$ is obtained by restricting
(\ref{R21nullplane}) to the $t=0$ $\R^2$-plane, which results in 
\be \label{R2line} 
\hat \om = -\sin \theta \; x + \cos \theta \; y. 
\ee
This is the equation for oriented lines in $\R^2.$  Hence, we have
that the
space of finite real null planes (in spacetime $\RP^3$ or $\CP^3$) is the space of
oriented lines in $\R^2,$ which is $S^1 \times \R.$ 
From (\ref{R2line}) we note that  
$(\hat \om, \theta)$ and $(-\hat \om,\; \theta + \pi)$ give the same unoriented
line.   This means that the two
orientations of a line correspond to a pair of null planes labelled by  
$(\hat \om, \pi_0, \pi_1) $ and $ (-\hat
\om, -\pi_1, \pi_0),$ or  $[Z_0,Z_1,Z_2,Z_3] $ and 
$ [Z_0,Z_1,Z_2,-Z_3].$

\vspace{0.5cm}

Now, let us consider non-real null planes, given by the points  
$[Z_\a] \in {\mc C}-{\mc C}_{\R},$ and again first look at $[Z_\a]$ with $Z_0 \ne 0.$ Such a null plane must have
either $z_1$ or $z_2$ non real, or both.  Writing 
$z_1=l+im$ and $z_2 = k+in,$ the plane equation  
\be \label{intersection}  P^0Z_0+P^1Z_1 + P^2Z_2 =0, \ee
 becomes
\[ P^0 = -(P^1l+P^2k) \qquad \mbox{and} \qquad P^1m+P^2n=0. \]
There are two cases.  If $m \ne 0,$ then
\[ P^1= - P^2\frac{n}{m} \qquad \mbox{and} \qquad  P^0 = P^2 \left( \frac{n}{m}l-k
\right),\]
and if $n \ne 0,$
\[ P^2= - P^1\frac{m}{n},  \qquad \mbox{and} \qquad P^0 = P^1 \left( \frac{m}{n}k-l
\right).\]
In other words, each non-real null plane with $Z_0 \ne 0$ intersects $\tau_\R$ in a
single point given by
\[ [\frac{n}{m}l-k, \; -\frac{n}{m}, \; 1] \qquad \mbox{and} \qquad
   [\frac{m}{n}k-l, \; 1, \;  -\frac{m}{n}]  \]
for $m \ne 0$ and $n \ne 0$ respectively.  Note that if $m \ne 0$ it
follows that
$P^2 \ne 0$ and if $n \ne 0$ then $P^1 \ne 0.$  

\vspace{0.5cm}

Now consider 
non-real null planes with $Z_0 =0.$  First, note that since ${Z_0 =0,}$
both $Z_1, Z_2$ must be non-zero, otherwise, for example if $Z_2 = 0$ the cone equation $(Z_1)^2
- (Z_3)^2 = 0$ implies that the plane labelled by $[0, Z_1, 0, Z_3]$
is a real null plane.  Now, let us write $Z_1 = L+iM, \; Z_2 = K+iN.$  Then
(\ref{intersection}) implies that 
\[ LP^1 +KP^2 = 0, \qquad MP^1 +NP^2 = 0.  \]
For a non-real null plane, at least one of $M$ or $N$ must be
non-zero.  Suppose $M \ne 0$ then $P^1 = -\frac{N}{M}P^2.$  There
are 2 cases.
 
\begin{enumerate}[(i)]
 \item $L \ne 0:$ \, Then $P^1 = - \frac{K}{L}P^2.$  Since either $N$ or $K$
must be non-zero, in a generic case where $\frac{N}{M} \ne
\frac{K}{L}$ we have $P^1 = 0 = P^2.$  Therefore the plane intersects
$\tau_\R$ at a single point $[P^0, P^1, P^2] = [1, 0,0].$  It can be
shown using the cone equation (\ref{homocone}) that, if  $\frac{N}{M} =
\frac{K}{L},$ then the null planes are real null planes. 
\item $L=0:$ \, Then $KP^2 = 0.$  If $K \ne 0$ we again have $P^1 = 0 =
P^2.$  On the other hand $K=0$ means $Z_1, Z_2$ are pure imaginary,
which implies that $Z_3$ is also pure imaginary.  Thus the plane is a
real  null plane. 
\end{enumerate}

If we suppose $L \ne 0$ at the beginning, interchanging the roles of $(L, M)$ and $(K, N)$ yields the same
result.  Hence we conclude that each non-real null plane intersects
$\tau_\R$ at a single point.

Therefore,  every null plane in $\ov{M_\C} = \CP^3$ intersects
$\tau_{\R}.$  \koniec

\noindent {\bf Remark.} 
The surjectivity of the restricted map (\ref{themap}) is due to the
fact that under the map (\ref{qdef2}) each point on $L_\infty$
corresponds to $\CP^2_\infty.$ 
 We have (\ref{qdef2}) essentially because we take the
correspondence space $\h {\mc F}$ of the fibration to be the blow-up of the variety $\h
f$ along $\CP^2_\infty \times \{{\bf z_0}\}.$   

Recall Lemma \ref{Linftytocp1} which states that each point on $L_\infty$ corresponds to a $\CP^1$ line in
$\CP^2_\infty$ under (\ref{tworoots}).  We note here that not every point in
$L_\infty$ gives a $\CP^1$ that 
intersects $\tau_\R.$  Consider equation (\ref{tworoots}) with $P^3 = 0$ for the
intersection of such a $\CP^1$ with the constant time slice $\tau:$  
\be \label{pointeq} -2 \ll P^1 + (1-\ll^2)P^2 =0.\ee
Since the coefficients  of $P^1$ and $P^2$ cannot be zero at the same time, we have one
degree of freedom in $(P^1, P^2).$  Hence, the $\CP^1$ intersects
$\tau$ at a single point.  For example if $\ll
\ne 0,$ the intersection point is given by
\be \label{point}
 [P^1, P^2] = [\frac{(1-\ll^2)}{2\ll}, 1].
\ee
Note that the map is $2:1$ as  $\ll$  and $-\frac{1}{\ll}$ give
the same point in $\tau.$  

Now assume that $[P^1,P^2] \in \tau_{\R}.$  From
(\ref{point}), we need $\frac{(1-\ll^2)}{2\ll}$ to be real.  Writing
$\ll =l+im,$ the imaginary part of $\frac{(1-\ll^2)}{2\ll}$ is
$\frac{-m(1+l^2+m^2)}{l^2+m^2}.$  This vanishes if and only if $m=0.$
Therefore, we conclude that a point $\ll$ in $L_\infty$
gives rise to a $\CP^1$ in $\CP^2_\infty$ which intersects
$\tau_{\R}$ if and only if $\ll \in \RP^1 \subset L_\infty.$  Then the
intersection is a single point determined by (\ref{pointeq}). 

\vspace{0.5cm }

\section{Involution maps}   \label{appe}
\setcounter{equation}{0}

\vspace{0.3cm}

\noindent {\bf \large Time reversal}

In the noncompact spacetime $M_\C = \C^3$ with coordinates $(x,y,t),$ we
define the time reversal map as usual as 
\be \label{tspace} \sm: (x,y,t) \longmapsto (x,y,-t). \ee
The fixed points of (\ref{tspace}) are of course those with $t=0.$
The map (\ref{tspace}) induces a holomorphic involution on $T\bP^1$  via the null
plane equation
\be \label{mapnullplane} \hat \om = 2x\pi_0\pi_1 + y (\pi_0^2 -\pi_1^2) + t(\pi_0^2
+\pi_1^2).\ee
Under (\ref{tspace}), equation (\ref{mapnullplane}) becomes
\be \label{timagenull} \hat \om = 2x\pi_0\pi_1 + y (\pi_0^2 -\pi_1^2) - t(\pi_0^2
+\pi_1^2).\ee
We now want to define a map $\sm$ (keeping
the same name) acting on a point $(\hat \om, \pi_A)  \in T\bP^1$ such that
the  image $(\hat \om', \pi_A')$  corresponds to the null plane defined
by (\ref{timagenull}).  Multiplying (\ref{timagenull}) by $-1$ on both
sides does not change the
plane and we have  
\[ \sm: (\hat \om, \pi_0, \pi_1) \longmapsto (\hat \om' = -\hat \om,\;
\pi_0' = -\pi_1, \;
\pi_1'=\pi_0).\]
We could equally well define the map $\sm$ with
$\pi_A'=(\pi_1, -\pi_0),$ but since $\pi_A$ are homogeneous
coordinates of $\CP^1,$ the choice does not matter.  
In the inhomogeneous coordinates  $\dsl (\om = \frac{\h \om}{\pi_1^2}, \; \ll =
\frac{\pi_0}{\pi_1}) $ we have 
\[\sm: (\om, \ll) \longmapsto (-\t \om , -\t \ll), \] 
where we recall that $\dsl \t \om = \frac{\om}{\ll^2}$ and $\dsl \t \ll =
\frac{1}{\ll}$ in the overlap.   It is immediate that a holomorphic
section (\ref{mapnullplane}) labelled by $(x,y,t)$ is  preserved by
$\sm$ if and only if $t=0.$ \\

For the compact case, we want to extend the map (\ref{tspace}) to
$\ov {M_\C} = \CP^3$ and define the corresponding involutions on the cone
${\mc C} \subset {\CP^3}^*$ and its blow-up $\wt {\mc C} \cong  \ov 
{T\bP^1}.$  Recall our convention that the extension of $t=0$ \, $\C^2$-plane to $\ov{M_\C}$ is
 $ \tau = \CP^2$ defined by $P^3=0,$ and $(x, y, t)$ is given by
(\ref{xytcoord}).  
Then the extension of (\ref{tspace}) to
 $\ov {M_\C}$ is
\[ \sm: [P^0,P^1,P^2,P^3] \longmapsto [P^0,P^1,P^2,-P^3].\]
This induces a map $\sm$ on ${\CP^3}^*$ via (\ref{homoplane}), given by
\be \label{tmap} \sm: [Z_0,Z_1,Z_2,Z_3] \longmapsto [Z_0,Z_1,Z_2,-Z_3],\ee
and hence a map on the cone ${\mc C}.$ 

By generalising the discussion in \ref{appd}, we can 
show that each null plane in $\ov {M_\C}$ intersects $\tau$ in an $\CP^1.$  Moreover, two null planes
labelled by
\[ [Z_0, \; Z_1, \; Z_2, \; +\sqrt{Z_1^2+Z_2^2}] \qquad \mbox{and} \qquad 
[Z_0,\; Z_1, \; Z_2, \;  -\sqrt{Z_1^2+Z_2^2}] \]
have the same intersection line.  Hence, geometrically $\sm$
interchanges the two members of such pair.  
The fixed points of the map are the planes which do not form a
pair.   A special case is the vertex of the cone ${\bf z_0}$ with
$\CP^1_\infty$ intersection.  The rest are those with 
$Z_2 = \pm i Z_1.$  These are the points $[\frac{Z_0}{Z_1}, 1, \pm i,
  0] \in {\mc C},$ and there are two $\C$-worth sets of these
points.  Choosing new representatives as $[\pm 2i\frac{Z_0}{Z_1}, \pm2i, 2,
  0],$ we see that these are the fibres $\ll = \pm i$ of $T\bP^1.$

We can extend the map (\ref{tmap}) to the compactified twistor space
$\ov {T\bP^1},$ which is biholomorphic to $\wt {\mc C},$ by demanding that it gives back (\ref{tmap}) under
the projection ${\pi: \wt {\mc C} \ra {\mc C}.}$  Locally, say in
the blow-up $\wt {\mc C}_U \subset U \times \CP^2  $ of the patch $U$
with $Z_0 \ne 0,$ the map is defined by its action on $U \times \CP^2$
as 
\[\sm: (z_1,z_2,z_3) \times [l_1,l_2,l_3] \longmapsto  
(z_1,z_2,-z_3) \times [l_1,l_2,-l_3],\]
where $z_i = \frac{Z_i}{Z_0}$ are local coordinates on
$U.$  Note that the blow-up vertex $L_\infty \subset \ov {T\bP^1}$ 
is fixed by $\sm,$ although it is not the set of fixed points.

\vspace{1cm}

\noindent{\bf \large Reality condition}

\vspace{0.3cm}

Define a  map $\vp: M_\C \ra M_\C$ by complex conjugation
\be \label{rspace} \vp: (x,y,t) \lmt (\bar x, \bar y, \bar t).\ee  
The set of fixed
points of (\ref{rspace}) is the real slice $\R^{2,1} \subset M_\C.$
Now, considering the null plane equation  (\ref{mapnullplane}) one
sees that the map $\vp$
induces an anti-holomorphic involution on $T\bP^1$ which maps each point to its complex
conjugate 
\be \label{rtwist}  \vp: (\hat \om, \pi_A)  \lmt (\b {\hat \om}, \bar \pi_A). \ee
The fixed points of (\ref{rtwist}) corresponds to real null planes
discussed in Section \ref{sec:restcorres}.  Hence the set of fixed point
is $TS^1 \subset T\bP^1.$ 

There are unique extensions of (\ref{rtwist}) to $\ov {M_\C} = \CP^3$ and ${\CP^3}^*,$ sending
\[ [P^0,P^1,P^2,P^3] \lmt [\bar P^0, \bar P^1, \bar P^2, \bar P^3]
\quad \mbox{and} \quad [Z_0,Z_1,Z_2,Z_3] \lmt [\bar Z_0, \bar Z_1,
  \bar Z_2, \bar Z_3],\]
respectively.  The cone ${\mc C} \subset {\CP^3}^*$ is preserved by
  the map, with the vertex ${\bf z_0}$ being another fixed point in addition
  to the set $TS^1.$  

The extension to the blow-up $\wt U$ of the neighbourhood $U$ around
${\bf z_0}$ is obtained similarly to the case of the time reversal, where the map is
given locally by the complex conjugation of the coordinates of $\wt U.$  The involution
$\vp$ maps the blow-up vertex 
\[ L_\infty =\{ [l_i] \in \CP^2 :
l_1^2+l_2^2-l_3^2=0 \}\] 
to itself, and the fixed points  are those
with $[l_i] \in \RP^2 \subset \CP^2.$  In the coordinate $\ll \in
L_\infty,$ these are the points with real $\ll.$

Finally, $\vp$ preserves the sections in $\ov {T\bP^1}$
corresponding to $[P^\a] \in \RP^3.$   For finite-point sections it follows
readily from (\ref{section}).  For the sections corresponding to the
points at infinity, the pairs
of lines of constant $\ll$ are determined by (\ref{tworoots}).  We see
that for $\{P^\a\}$ real, the two roots can either be both real or complex
conjugates, and thus the pairs of lines are preserved by $\vp.$


\end{document}